\documentclass[preprint,3p,times,onecolumn]{elsarticle}
\usepackage{amsmath}
\usepackage{epsfig}
\usepackage{amssymb,amsthm,graphicx}
\usepackage{tabularx}
\usepackage{multirow}
\usepackage{tabularx}
\usepackage{float}

\newcommand{\bea}{\begin{eqnarray}}
\newcommand{\eea}{\end{eqnarray}}
\newcommand{\bes}{\begin{subequations}}
\newcommand{\ees}{\end{subequations}}

\newcommand{\sn}{{\rm sn}}

\newcommand{\dn}{{\rm dn}}
\newcommand{\cn}{{\rm cn}}
\newcommand{\sech}{{\rm sech}}

\journal{Physics Letters A}

\begin{document}

\begin{frontmatter}
\title{Elliptic Waves in Two Component Long Wave--Short Wave Resonance Interaction System in One and Two dimensions}

\author{Avinash Khare} \ead{khare@iiserpune.ac.in}
\address{Raja Ramanna Fellow, Indian Institute of Science Education and Research, Pune 411021, India}
\author{T. Kanna\corref{tk}}\ead{kanna\_phy@bhc.edu.in}
\author{K. Tamilselvan} \ead{tamsel786@gmail.com}
\address{Post Graduate and Research Department of Physics, Bishop Heber College, Tiruchirapalli 620 017, Tamil Nadu, India}
\cortext[tk]{Corresponding author}

\begin{abstract}
We consider (2+1) and (1+1) dimensional long-wave short-wave resonance interaction systems. We construct an extensive set of exact periodic solutions of these systems in terms of Lam\'e polynomials of order one and two. The periodic solutions are classified into three categories as similar, mixed, superposed elliptic solutions. We also discuss the hyperbolic solutions as limiting cases.
\end{abstract}

\end{frontmatter}

\section{Introduction}
The study of nonlinear waves is of broad scientific interest \cite{whitham}. Nonlinear waves in multi component long-wave-short-wave resonant interaction (LSRI) system have received significant attention in recent years. Here nonlinear resonance interaction between a low frequency long-wave (LW) and multiple high frequency short-waves (SWs) takes place when the phase velocity (say $v_p$) of the former exactly or approximately matches with the group velocity (say $v_g$) of the short-waves, that is, $v_p$ $\simeq$ $v_g$.  This LSRI phenomenon has a wide range of applications ranging from water waves to nonlinear optics which also include bio-physics and plasma physics. In the SW components the soliton is formed due to a delicate balance between the dispersion and the nonlinear interaction of LW with the SWs while in the LW component, the soliton formation is determined solely by the self-interaction of short wave-packets.

The pioneering work of LSRI system was done by Zakharov \cite{zak}. Later on, the general Zakharov equations in one dimension have been reduced to the integrable Yajima-Oikawa system in Ref.~\citep{yo}. At the same time, independently Benney has derived the model equation for the interaction of short wind-driven capillary gravity wave in deep water \cite{benny}. The experimental study of LSRI in a three layer fluid was carried out by Kopp and Redekopp \cite{kopp}. Then, in a physical set up of two layer fluid model the one and two dimensional LSRI systems have been derived and bright and dark soliton solutions have also been obtained in Refs. \cite{fun,oik}.

Recently, Kanna \textit{et al}. have  shown that the following (1+1) dimensional (i.e. one time and one space dimensions) LSRI system \cite{tk_pre}
\bes\label{1}\bea
&iS_{j,t}+\delta S_{j,xx}+ L S_{j}=0,\quad \quad j=1,2,\label{1a}\\\label{1b}
&L_t =  2\displaystyle\sum_{j=1}^{2}c_{j}|S_j|^2_{x},
\eea
\ees
 can be deduced from a set of three coupled nonlinear Schr{\"o}dinger equations governing the propagation of three optical fields in a triple mode optical fiber, by applying the asymptotic reduction procedure. In eq. (1), $S_{j}$ and $L$, respectively, indicate $j^{th}$ short wave and (one) long wave, $t$ and $x$ represent the partial derivatives with respect to evolutional and spatial coordinates, respectively, and the nonlinearity coefficients $c_{j}, j=1,2,$ are arbitrary real parameters. Here $\delta=\pm1$ and for $\delta=1$ the above system (\ref{1}) is nothing but the two component Yajima-Oikawa (YO) system.  Eq. (\ref{1}) also appears in the study of interaction of quasi resonant two frequency short wave pulses with a long wave \cite{sazo}. Such multi-component YO system also has been derived in the context of multiple component magnon-phonon system \cite{myrz}.  In Ref. \cite{tk_pre}, we have obtained the bright n-soliton solution of the above system (1) and have revealed the  fact that the bright solitons can undergo two types of  fascinating energy sharing collisions. Here the presence of the long wave induces nonlinear interaction between two SWs which leads to nontrivial collision behaviour. The rogue waves of LSRI system with $j=1$ (one SW and LW components) and $\delta=1$ have been reported in Ref. \cite{chen}. \\
The two dimensional multi-component LSRI system has also received equally good attention as that of their one-dimensional counterpart. Particularly the following two component analogue of the (2+1) dimensional (i.e. two spatial coordinates $x$, $y$ and one time coordinate) LSRI system
\bes \label{2}\bea
&i[S_{j,t}+\varepsilon_{j} S_{j,y}]+\delta~S_{j,xx}+ L S_{j}  =0, \quad j=1,2,\label{2a}\\
&L_t = 2 \displaystyle\sum_{j=1}^{2}c_{j}|S_j|^2_{x},\label{2b}
\eea \ees
where the subscripts $x$ and $y$ represent the partial derivatives with respect to spatial coordinates and $t$ represents temporal coordinate, the nonlinearity coefficients $c_{1}$ and $c_{2}$ and the co-efficient $\varepsilon_{j}$, $j=1,2$, are real arbitrary parameters and $\delta=\pm 1$. Eq. (2) has been derived as the governing equation for the interaction of three nonlinear dispersive waves in optical fiber or in photorefractive medium  by applying a reductive perturbation method \cite{ohta}. In the above system, two SWs propagate in anomalous dispersion regime and the real long wave propagates in the normal dispersion regime. In a recent work \cite{tk_lanl},  Kanna \textit{et al}. have generalized the approach of \cite{ohta} and derived a $M$-component LSRI system as the propagation equation for multiple dispersive waves (say $(M+1)$ waves ) in a weak Kerr type nonlinear medium in the small amplitude limit. To get further physical insight into the above system (\ref{2}), we would like to point out that the one component ($j=1$) version of Eq. (\ref{2}) can be derived from the governing equation for two-dimensional two wave interaction \cite{onor,shukla} by following the approach of \cite{ohta}. Thus system (\ref{2}) is a three- wave generalization of two wave system in (2+1) dimensions. From a mathematical perspective, the soliton solutions of system (2), with $c_{1}=c_{2}=1$ are constructed in Refs. \cite{ohta, tk_jpa_lsri}. Particularly, in Ref. \cite{tk_jpa_lsri} it has been shown that the bright solitons exhibit interesting energy sharing collisions characterized by intensity (energy) redistribution, amplitude dependent phase-shifts and change in relative separation distances.

Periodic nonlinear waves can also arise in real physical systems. For example, generation of ultrashort pulse-train by using nonlinear transform of a twin frequency signal is one such real system arising in nonlinear optics (\cite{chow1} and references there in). Thus to describe real situations one may need special type of periodic solutions. Several periodic solutions of integrable and nonintegrable mullticomponent nonlinear Schr{\"o}dinger equations with focusing, defocusing and mixed type nonlinear interactions have been obtained in Refs. \cite{chow1}-\cite{wright}, in terms of Jacobi elliptic functions. So far, such elliptic wave solutions have not been constructed for the  one- and two- dimensional two component LSRI systems (1) and (2) as these integrable systems have been reported recently. This paper is aimed at constructing different families of elliptic wave solutions of (1) and (2) in a systematic way.

The organization of the paper is as follows. In section 2, the elliptic  wave solutions of the $(2+1)$ dimensional two component LSRI system (2) are obtained in terms of Lam{\'e} polynomials of orders one and two. Similar solutions of $(1+1)$ dimensional two component LSRI system (1) are dicussed in section 3. Finally, conclusions are drawn in the last section.

\section{Jacobi Elliptic function solutions of the (2+1)-dimensional two component LSRI system}
We start with the $(2+1)$ dimensional LSRI system (\ref{2}). To construct the Jacobi elliptic solutions of eq. (\ref{2a}) we choose the travelling wave ansatz
\bea \label{3}
S_{j}(x,y,t)=f_{j}[\beta(x-vt-w y+\delta_{0})]e^{-i(\omega_{j}t+\nu_{j}y-k_{j}x+\delta_{j})},&&j=1,2.
\eea
 Here $f_{j}$ are real functions of $x$, $y$ and $t$; $\beta, \delta_{0}$ and $\delta_{1,2}$  are real constants, $\omega_{j}$ is the frequency of the $j^{th}$ SW component, $k_{j}$ is the wave number, $v$ is the velocity, $w$ and $\nu_{j}$ are real parameters. Note that both the short waves are travelling with the same velocity. Inserting the above ansatz (\ref{3}) into Eq. (\ref{2b}), we obtain the LW component as
\bea\label{4}
L=-\frac{2}{v}(c_{1}|S_{1}|^{2}+c_{2}|S_{2}|^{2}).
\eea
Following this, by substituting the ansatz (\ref{3}) into (\ref{2}) and also by using (\ref{4}), we get a set of complex equations. On equating the real and imaginary parts, we respectively obtain
\bes\label{5}\bea
\frac{d^2 f_j}{du^2}+ \left[ \frac{\delta (\omega_{j}+\varepsilon_{j} \nu_{j})-k_{j }^{2}}{\beta^2}-\frac{2 \delta}{v \beta^2}\left(c_1 f_1^2+c_2 f_2^2\right)\right]f_j=0,\;\;\;j=1,2,\label{5a}\\ \label{5b}
v+\varepsilon_{j}w=2\delta k_{j},\qquad \qquad \qquad \qquad \qquad \qquad
\eea\ees
where $u=\beta(x-v t-w y+\delta_{0})$. At a first look it might seem that (\ref{5}) is similar to the coupled nonlinear Schr\"odinger system given in \cite{hioe2}. But a careful analysis shows that they are essentially different. This is due to the presence of the long wave component $L$. In fact, here the solution parameters $\beta$ and the velocity $v$ explicitly appear before the non-linear term $(c_{1} f_{1}^{2}+c_{2} f_{2}^{2})$ . This makes the present system different from that of \cite{hioe2}. Particularly, this $v$ determines the nature of the solution, i.e., whether the solution is singular or not, as will be shown later. Thus the results presented here are distinct from those given in \cite{hioe2}, though the elliptic function solutions take standard Lam\'e function profiles as will be demonstrated below. These solutions can be viewed as velocity locked solutions.\\
Next, we assume the Lam\'e function ansatz for $f_{j}$, that is,
\bea\label{6}
f_j=\rho_j \psi_{j}^{(l)},   \quad \quad l,j=1,2,
\eea
where $\psi_{j}^{(l)}$ can be anyone of the three first order Lam\'e polynomials for $ l=1$ and for $l=2$, it can be any one of the five second order Lam\'e polynomials and satisfy the Lam\'e equation \cite{whit},
\bea\label{7}
\frac{d^{2}\psi_{j}^{(l)}}{du^{2}}+[\lambda_{j}^{(l)} - l(l+1)m \sn^{2}(u,m)]\psi_{j}^{(l)}=0,
\eea
where $m$ ($0\leq m \leq 1$) is the modulus parameter of the Jacobi elliptic function  $\sn(u,m)$,~~$l~(=1,2)$ represents the order of the Lam\'e polynomial $\psi_{j}^{(l)}$ and $\lambda_{j}^{(l)}$ is the corresponding eigenvalue. Thus we will have two distinct families of solutions corresponding to the Lam\'e polynomials of order 1 ($l=1$) and of order 2 ($l=2$). First, we present and discuss periodic solutions in terms of  Lam\'e polynomials of order one and then we present the second order solutions.
%%%%%%%%%%%%%%%%%%%%

\subsection{Solutions in terms of Lam\'e polynomals of order 1}
The two component LSRI system (\ref{2}) admits seven distinct periodic solutions in terms of Lam\'e polynomials of order 1. These first order solutions of Eq. (\ref{2}) corresponding to $l=1$ can be expressed in terms of Jacobi elliptic functions \cite{abrow}. We classify these solutions as similar, mixed and superposed elliptic solutions. By similar we mean same kind of standard elliptic function profile for both the short wave components. For the mixed elliptic solutions, the two short wave components take distinct elliptic function profiles. The superposed solutions are special and they are constructed by superposition of two elliptic functions. Apart from these elliptic functions, we also discuss the hyperbolic (soliton/solitary wave) solutions by fixing the modulus parameter $m$ as one.
\subsubsection{Similar elliptic solutions}
The similar elliptic solutions are listed in Table 1.
\renewcommand{\floatpagefraction}{0.7}
\renewcommand{\arraystretch}{0.15}% Tighter
\begin{table}[H]
\caption {Similar elliptic solutions of order 1}
\centering
\begin{tabular}{|c|c|c|c|c|}
  \hline
    Similar&$f_{1}(u)$ & $f_{2}(u)$ & Constraints on & Long wave\\ \cline{2-3}
   elliptic &$S_{1}=f_{1}$&$S_{2}=f_{2}$&parameters& component\\
   solutions&$ \times e^{-i(\omega_{1}t+\nu_{1} y-k_{1}x+\delta_{1})}$&$ \times e^{-i(\omega_{2}t+\nu_{2} y-k_{2}x+\delta_{2})}$&~&~\\\hline
    \multirow{4}{1cm}{(1)} & \multirow{4}{2cm}{$A~\textit{\dn (u,m)}$}& \multirow{4}{2cm}{$B~\textit{\dn (u,m)}$}
          & $v + \varepsilon_{1} w =2\delta k_{1},~v +\varepsilon_{2} w = 2\delta k_{2}$,&$\left(2 \delta \beta^2\right)\dn^2(u,m) $ \\
        & & & $\beta^2 = -\delta\frac{\left(c_1 A^2 +c_2 B^2\right)}{v}>0$,&~ \\
        & & & $\omega_{1}+\varepsilon_{1} \nu_{1}-\delta\left(k_{1}^2-(2-m) \beta^2\right)=0$,&~ \\
        & & & $\omega_{2}+\varepsilon_{2} \nu_{2}-\delta\left(k_{2}^2-(2-m) \beta^2\right) =0$.&~ \\
  \hline
    \multirow{4}{1cm}{(2)} & \multirow{4}{2cm}{$A~\sqrt{m}~\textit{\cn (u,m)}$}& \multirow{4}{2cm}{$B~\sqrt{m}~\textit{\cn (u,m)}$}
          & $v + \varepsilon_{1} w =2\delta k_{1},~v + \varepsilon_{2} w =2\delta k_{2}$,&$\left(2 \delta m\beta^2\right)\cn^2(u,m)$\\
        & & & $\beta^2 = -\delta\frac{\left(c_1 A^2 +c_2 B^2\right)}{v}>0$,&~\\
        & & & $\omega_{1}+\varepsilon_{1} \nu_{1}-\delta\left (k_{1}^2-(2m-1) \beta^2\right) =0$,&~ \\
        & & & $\omega_{2}+\varepsilon_{2} \nu_{2}-\delta\left(k_{2}^{2}-(2m-1) \beta^2\right) =0$.&~ \\
    \hline
\multirow{4}{1cm}{(3)} & \multirow{4}{2cm}{$A~\sqrt{m}~\textit{\sn (u,m)}$}& \multirow{4}{2cm}{$B~\sqrt{m}~\textit{\sn (u,m)}$}
          & $v+ \varepsilon_{1} w =2\delta k_{1},~v + \varepsilon_{2} w=2\delta k_{2}$,&$\left(-2 \delta m\beta^2\right)\sn^2(u,m)$\\
        & & & $\beta^2 = \delta\frac{\left(c_1 A^2 +c_2 B^2\right)}{v}>0$,&~ \\
        & & & $\omega_{1}+\varepsilon_{1} \nu_{1} -\delta\left(k_{1}^2+ (1+m)\beta^2\right) =0$,&~ \\
        & & & $\omega_{2}+\varepsilon_{2} \nu_{2} -\delta \left(k_{2}^2+(1+m) \beta^2\right) =0$.&~ \\\hline
        \end{tabular}
   \end{table}
In Table 1, $(u,m)=(\beta(x-vt-wy+\delta_{0}),m)$. The long wave solutions are also given in the last column of Table 1. Each of the above solutions contains fifteen real parameters along with five constraints. Note that the period of second and third solution is twice that of first solution. The amplitudes of solution (1) for the two SW components are A and B while that of solutions (2) and (3) are $A\sqrt{m}$ and $B\sqrt{m}$. The distinction of the present solutions from the solutions of coupled nonlinear Schr\"odinger equations \cite{hioe2} mainly lies in the constraint conditions which significantly alters the nature of the solutions. Particularly, to get regular solution, we require $\beta^{2}$ to be positive. Once the signs of $c_{1}$ and $c_{2}$ as well as $\delta$ are fixed this can be achieved by suitably choosing the sign of $v$. To get insight into another physical aspect of the present solutions, let us consider the constraint expression for $\beta^2$ in solution (1) with $c_1=c_2=1$, for simplicity. Then one can find $A^2+B^2=-\frac{\beta^2 v}{\delta}$, $\delta = \pm 1$.  This indicates that for a given value of $\frac{\beta^{2}v}{\delta}$ any change in intensity (square of the amplitude) in one SW component will in turn influence the intensity of the other. Thus the solution displays intensity exchange. Ultimately, intensity of a particular SW component can be enhanced by suppressing that of the other component. Same conclusion can be drawn for the solutions (2) and (3) too. Such kind of intensity sharing depending upon the velocity is a special feature of the present two-component LSRI system.
\subsubsection{Mixed elliptic solutions}
The mixed elliptic solutions admitting different profiles of elliptic wave trains in the the short wave components $S_{1}$ and $S_{2}$ of Eq. (\ref{2}) are tabulated below.
\renewcommand{\floatpagefraction}{0.7}
\renewcommand{\arraystretch}{0.05}% Tighter
\begin{table}[H]
\caption {Mixed elliptic solutions of order 1}
\centering
\begin{tabular}{|c|c|c|c|c|}
  \hline
    Mixed &$f_{1}(u)$& $f_{2}(u)$ & Constraints on & Long wave\\\cline{2-3}
   elliptic&$S_{1}=f_{1}$&$S_{2}=f_{2}$&parameters& component\\
     solutions&$\times e^{-i(\omega_{1}t+\nu_{1} y-k_{1}x+\delta_{1})}$&$\times e^{-i(\omega_{2}t+\nu_{2} y-k_{2}x+\delta_{2})}$&~&~\\\hline
    \multirow{4}{0.5cm}{(1)} & \multirow{4}{2.5cm}{$A~\textit{\dn (u,m)}$}& \multirow{4}{2.5cm}{$B~\sqrt{m}~\textit{\sn (u,m)}$}
          & $v+ \varepsilon_{1} w =2\delta k_{1}$,&~$-2~[\left(\frac{c_{2}}{v}\right)~B^{2}$ \\
        & & & $v+ \varepsilon_{2} w =2\delta k_{2}$&~\\
        & & & $ \beta^2 = \delta\frac{\left(c_2 B^2-c_1 A^2\right)}{v}>0$,&$-\delta \beta^{2}\dn^{2}(u,m)]$ \\
        & & & $\omega_{1} +\varepsilon_{1} \nu_{1}-\delta\left(k_{1}^2 +m \beta^2\right)-\frac{2}{v} c_1 A^2 = 0$,&~ \\
        & & & $\omega_{2} +\varepsilon_{2} \nu_{2}-$&~ \\
        & & &$\delta\left(k_{2}^2 +(1+m) \beta^2\right)-\frac{2}{v} c_1 A^2 = 0$.&~\\
    \hline
    \end{tabular}
     \end{table}
\renewcommand{\floatpagefraction}{0.7}
\renewcommand{\arraystretch}{0.10}% Tighter
\begin{table}[H]
\centering
    \begin{tabular}{|c|c|c|c|c|}
    \hline
    \multirow{4}{0.5cm}{(2)} & \multirow{4}{2.5cm}{$A~\sqrt{m}~\textit{\cn (u,m)}$}& \multirow{4}{2.5cm}{$B~\sqrt{m}~\textit{\sn(u,m)}$}
          & $v+ \varepsilon_{1} w =2\delta k_{1}$,&~$-2[\left(\frac{c_{2}}{v}\right)~m~B^{2}$\\
        & & & $v+ \varepsilon_{2} w =2\delta k_{2}$,&~$-m \delta \beta^{2}\cn^{2}(u,m)]$\\
        & & & $~~\beta^2 = \delta\frac{\left(c_2 B^2 -c_1 A^2\right)}{v}>0$,&\\
        & & & $\omega_{1} +\varepsilon_{1} \nu_{1}$&~ \\
        & & & $-\delta\left(k_{1}^2 + \beta^2\right)-\frac{2}{v} m c_1 A^2 = 0$,&~\\
        & & & $\omega_{2} +\varepsilon_{2} \nu_{2}$&~ \\
        & & &$-\delta\left(k_{2}^2 +(1+m) \beta^2\right)-\frac{2}{v} m c_1 A^2 =0$.&~\\
  \hline
    \multirow{4}{0.5cm}{(3)} & \multirow{4}{2.5cm}{$A~\textit{\dn (u,m)}$}& \multirow{4}{2.5cm}{$B~\sqrt{m}~\textit{\cn (u,m)}$}
          & $v+ \varepsilon_{1} w =2\delta k_{1}$,& ~\\
        & & & $v+ \varepsilon_{2} w =2\delta k_{2}$,&$2~[\left(\frac{(1-m)}{v}\right)~c_{2}~B^{2}$\\
        & & & $\beta^2 = -\delta \frac{\left(c_1 A^2 +c_2 B^2\right)}{v}>0$,&$-\delta \beta^{2}\dn^{2}(u,m)]$ \\
        & & & $\omega_{1}+\varepsilon_{1} \nu_{1}$&~ \\
        & & & $-\delta\left(k_{1}^2-m \beta^2\right)+\frac{2}{v} (1-m) c_1 A^2 =0$,&~\\
        & & & $\omega_{2}+\varepsilon_{2} \nu_{2}$&~ \\
        & & &$-\delta \left(k_{2}^2- (1-2m)\beta^2\right)+\frac{2}{v} (1-m) c_1 A^2 =0$.&~\\
    \hline
    \end{tabular}
   \end{table}
In the above table, $(u,m)=(\beta(x-v t-wy+\delta_{0}),m)$. Here the solutions are characterized by fifteen real parameters along with five constraints. As in the similar solutions the intensity sharing property is exhibited by the above solutions. This is evident from the expression for $\beta^{2}$ in the column "constraint on parameter" in Table 2.
\subsubsection{Superposed elliptic solution}
 The two component LSRI system (\ref{2}) admits the following superposed elliptic solution \cite{khare}:
\bes \label{8}\bea
S_{1}(x,t)&=&\left[\frac{1}{2}(A~\textit{\dn(u,m)}+D~\sqrt{m}\textit{\cn (u,m)})\right]e^{-i(\omega_{1}t+\nu_{1} y-k_{1}x+\delta_{1})},\label{8a}\\
S_{2}(x,t)&=&\left[\frac{1}{2}(B~\textit{\dn(u,m)}+E~\sqrt{m}\textit{\cn (u,m)})\right]e^{-i(\omega_{2}t+\nu_{2} y-k_{2}x+\delta_{2})},\label{8b}\\
L&=&\left(\frac{\delta \beta^{2}}{2}\right)(\textit{\dn(u,m)}~\pm ~\sqrt{m}~\textit{\cn(u,m)})^{2}.
\eea
Here the parametric restriction can be expressed as \\
\bea
v+ \varepsilon_{1} w =2\delta k_{1},~v+ \varepsilon_{2} w =2\delta k_{2},~D=\pm A,~E = \pm B,~\beta^2 = -\delta\frac{\left(c_1 A^2 +c_2 B^2\right)}{v}>0,\nonumber\\~\omega_{1}+\varepsilon_{1} \nu_{1}-\delta\left(k_{1}^2 -\frac{(1+m)}{2} \beta^2\right) =0,~
~\omega_{2}+\varepsilon_{2} \nu_{2}-\delta \left(k_{2}^2- \frac{(1+m)}{2} \beta^2\right)=0.
\eea \ees
Here the signs of $D$ and $E$ are correlated i.e. for $D=A~(D=-A)$, $E=B~(E=-B)$.
\subsubsection{Hyperbolic solution($m=1$)}
(a) In the limit $m=1$, mixed elliptic solutions (1) and (2) of Table 2 go over to the hyperbolic solution
\bes \label{9}\bea\label{9a}
S_{1}&=&\left[A~\sech(\beta(x-v t-w y+\delta_{0}))\right]~e^{-i(\omega_{1}t+\nu_{1}y-k_{1}x+\delta_{1})},\\\label{9b}
S_{2}&=&\left[B~\tanh(\beta(x-v t-w y+\delta_{0}))\right]~e^{-i(\omega_{2}t+\nu_{1}y-k_{2}x+\delta_{2})},\\
L&=&-2\left[\frac{c_{1} A^{2}}{v}+\delta \beta^{2}\tanh^{2}(\beta(x-vt-wy+\delta_{0}))\right],
\eea
with the parametric restriction\\
\bea
v+ \varepsilon_{1} w =2\delta k_{1},~v+ \varepsilon_{2} w =2\delta k_{2},~\beta^{2}=\delta\frac{\left(c_{2} B^{2}-c_{1}A^{2}\right)}{v}>0,\nonumber\\
\omega_{1}+\varepsilon_{1} \nu_{1}-\delta\left(k_{1}^{2}+\beta^{2}\right)-\frac{2}{v}c_{1}A^{2}=0, ~
\omega_{2}+\varepsilon_{2} \nu_{2}-\delta\left(k_{2}^{2}+2\beta^{2}\right)-\frac{2}{v}c_{1}A^{2}=0.
\eea \ees
Here the SW components $S_{1}$ and $S_{2}$ are comprised of bright and dark solitary waves,  respectively while the LW component is comprised of anti-dark (bright solitary wave with non-zero asymptotics) solitary wave. In Eq. (\ref{2}), the nonlinearity is said to be of focusing type, if the co-efficients ($c_{1}, c_{2}$) of nonlinearity and $\delta$ appearing before the dispersion term admit same sign. In that case, usually the system (\ref{2}) admits bright soliton solution. But here the nonlinear interaction of SWs with LW leads to the possibility of such mixed (bright-dark) solitary wave solution in system (\ref{2}) even with focusing type nonlinearities.  \\
(b) Further, the mixed elliptic solution (3) of Table 2, the similar elliptic solutions (1) and (2) given in Table 1 and superposed elliptic solution (\ref{8}) reduce to the hyperbolic (localized solitary wave/soliton) form given below:
\bes \label{10}\bea
S_{1}&=&\left[A~\sech(\beta(x-v t-w y+\delta_{0}))\right]~e^{-i(\omega_{1}t+\nu_{1} y-k_{1}x+\delta_{1})},\label{10a}\\
S_{2}&=&\left[B~\sech(\beta(x-v t-w y+\delta_{0}))\right]~e^{-i(\omega_{2}t+\nu_{2} y-k_{2}x+\delta_{2})},\label{10b}\\
L&=&\left(2 \delta \beta^{2}\right)\sech^{2}(\beta(x-vt-wy+\delta_{0})),
\eea
along with parametric restriction\\
\bea
v+ \varepsilon_{1} w =2\delta k_{1},~v+ \varepsilon_{2} w =2\delta k_{2},~\beta^{2}=-\delta\frac{\left(c_{1}A^{2}+c_{2}B^{2}\right)}{v}>0,\nonumber\\
\omega_{1}+\varepsilon_{1} \nu_{1}-\delta\left(k_{1}^{2}-\beta^{2}\right)=0,~
\omega_{2}+\varepsilon_{2} \nu_{2}-\delta\left(k_{2}^{2}-\beta^{2}\right)=0.
\eea \ees
Thus the SW and LW components admit bright-type solitary wave solutions. If the signs of ($c_{1}, c_{2}$) and $\delta$ are opposite then the nonlinearity is said to be of defocusing type. Normally, this nonlinearity will support dark or bright-dark solitons/solitary waves. From Eq.~(\ref{10}), we observe that even for this choice $\beta^{2}$ can be made positive by suitably choosing the sign of the velocity and the system can admit bright type solitary waves. This is a feature of two component LSRI system (\ref{2}).\\
(c) Finally, the similar elliptic solution (3) in Table 1 reduces to the hyperbolic solution with a kink-like form
\bes \label{11}\bea
S_{1}&=&\left[A~\tanh(\beta(x-v t-w y+\delta_{0}))\right]~e^{-i(\omega_{1}t+\nu_{1} y-k_{1}x+\delta_{1})},\label{11a}\\
S_{2}&=&\left[B~\tanh(\beta(x-v t-w y+\delta_{0}))\right]~e^{-i(\omega_{2}t+\nu_{2} y-k_{2}x+\delta_{2})},\label{11b}\\
L&=&-\left(2 \delta \beta^{2}\right) \tanh^{2}(\beta(x-vt-wy+\delta_{0}))\label{11c},
\eea
along with parametric restriction\\
\bea
v+ \varepsilon_{1} w =2\delta k_{1},~v+ \varepsilon_{2} w =2\delta k_{2},~\beta^{2}=\delta\frac{\left(c_{1} A^{2}+c_{2} B^{2}\right)}{v}>0,\nonumber\\
\omega_{1}+\varepsilon_{1} ~\nu_{1}-\delta\left(k_{1}^{2}+2\beta^{2}\right)=0,~
\omega_{2}+\varepsilon_{1}~\nu_{2}-\delta\left(k_{2}^{2}+2\beta^{2}\right)=0.
\eea \ees
Here the SW and LW components admit dark-type solitary wave solutions. It is interesting to notice that even for focusing nonlinearity ($c_{1},c_{2}$ and $\delta$ having same sign) the two components of LSRI system support dark-dark solitary waves through the nonlinear interaction of SWs with LW.
\renewcommand{\floatpagefraction}{0.7}
\begin{figure}[H]
\centering\includegraphics[height=4cm,width=5.8cm]{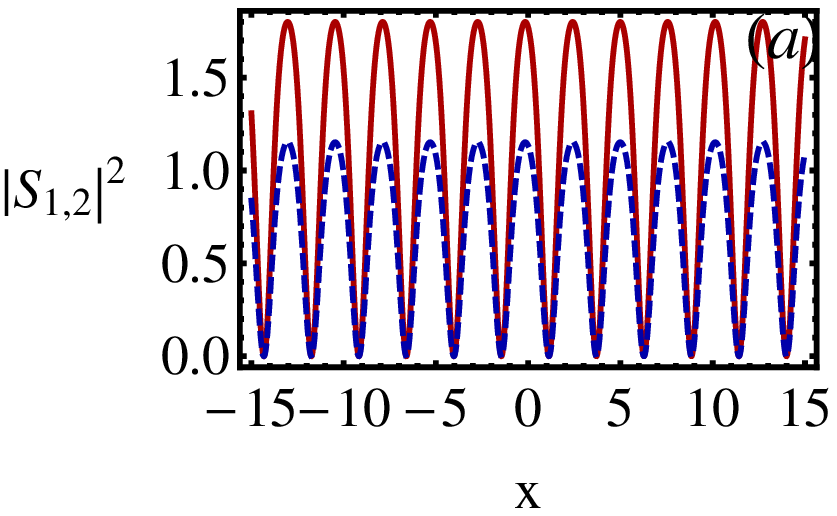}~~~\includegraphics[height=4cm,width=5.8cm]{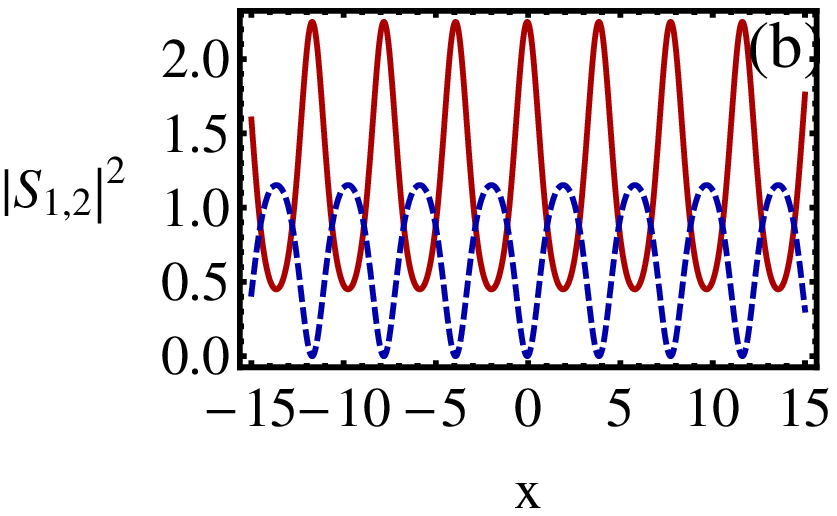}\\
\includegraphics[height=4cm,width=5.8cm]{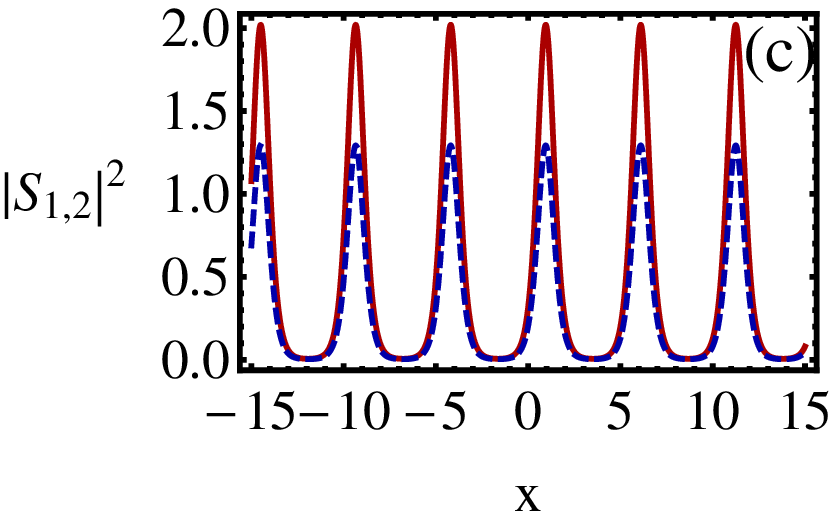}~~~\includegraphics[height=3.85cm,width=5.8cm]{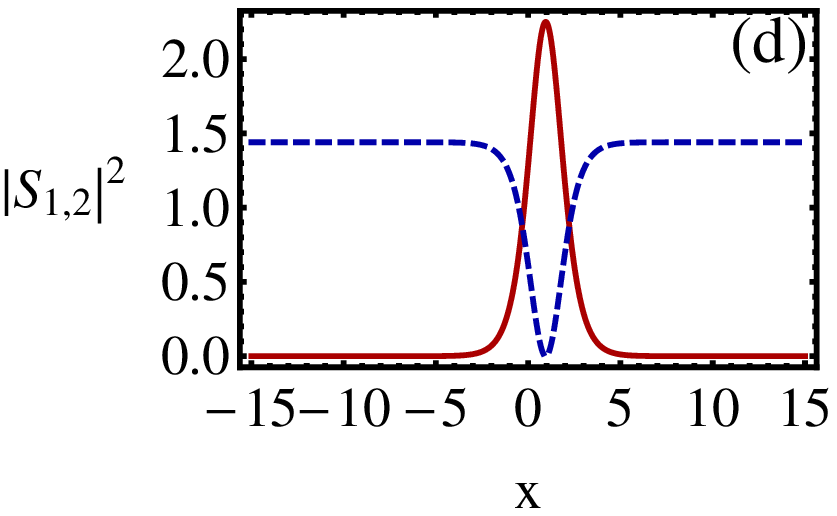}
\end{figure}
\begin{figure}[H]
\caption{Red and dashed blue plots correspond to first and second components of short wave, respectively. (a) intensity plot of similar elliptic solution (3) given in Table 1, for the velocity value  $v=-1.2$, $m=0.8$. (b) intensity plot of mixed elliptic solution  (1) given in Table 2, for the velocity value $v=0.2$, $m=0.8$. (c) superposed solution given by Eq. (\ref{8}) for the velocity value $v=1.2$, $m=0.8$ and for the condition $A=D$, $B=E$. (d) plot of regular  bright-solitary wave solution in $S_1$ component and dark-solitary wave solution in $S_2$ component given by Eqs. (\ref{9a}) and (\ref{9b}) for the velocity value $v=1.2$, $m=1$. In all the figures, $y=0.5$, $t=1$, the parameters $A=1.5$, $B=1.2$, $\nu_{1}=\nu_{2}=1.2$, $w=1.5$, $\delta=-1$, $\varepsilon_{1,2}=1$, $\delta_{0}=\delta_{1}=\delta_{2}=c_{1}=c_{2}=1$.}
\end{figure}
For illustrative purpose, we present some of the intensity plots of elliptic solutions of the short wave components in Fig.~1. Fig.~1(a) shows the periodic solution given by the similar solution (3) in Table 1. Fig.~1(b) shows the intensity plots of short waves given by mixed elliptic solution (1) in Table 2. It can be observed from Fig.~1(a) that the maximum of $S_{1}$ and $S_{2}$ are in-phase as they occur at the same value of $x$, while in the Fig.~1(b) they are out-of-phase as the intensity of one component reaches maximum when the other component takes minimum vale. So one can also refer to the corresponding elliptic function solutions as in-phase and out-of-phase solutions. Then, superposed periodic wave solutions for the short wave components ((\ref{8a}) and (\ref{8b})) are depicted in Fig.~1(c). These superposed solutions show that there is a significant increase in the pulse amplitude as compared with Fig.~1(a), even for same amplitudes A and B. Thus superposed elliptic solutions can be successfully employed for the generation of amplified pulse trains. Finally, in Fig.~1(d) we have displayed the hyperbolic (bright-dark soliton) solutions given by Eqs. (\ref{9a}) and (\ref{9b}). Here the first component is comprised of  bright solitary wave (soliton) and the second component of short wave is a dark solitary wave (soliton). These type of solitary waves arise with mixed boundary conditions $S_{1}\longrightarrow0$ and $S_{2}\longrightarrow$ constant as $x\longrightarrow \pm \infty$.
\renewcommand{\floatpagefraction}{0.7}
\begin{figure}[H]
\centering\includegraphics[height=4.1cm,width=5.8cm]{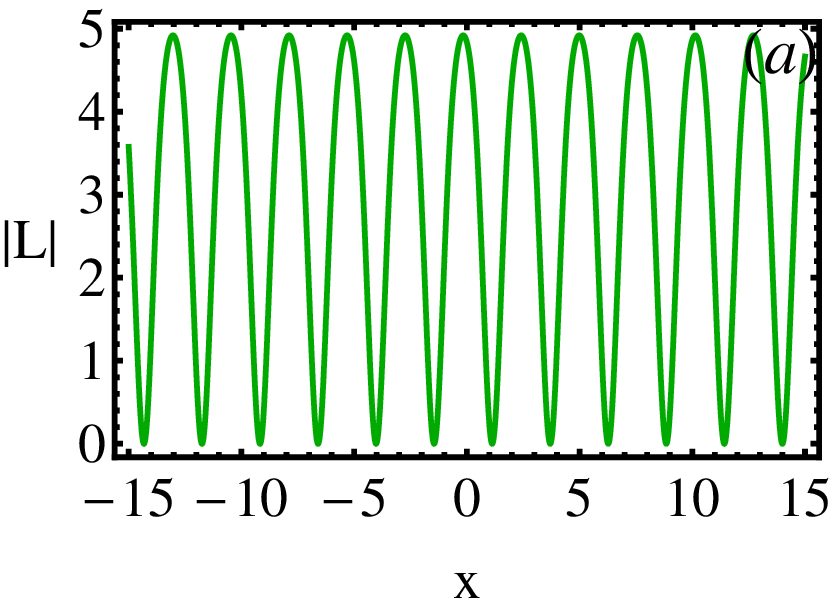}~~~\includegraphics[height=4cm,width=5.8cm]{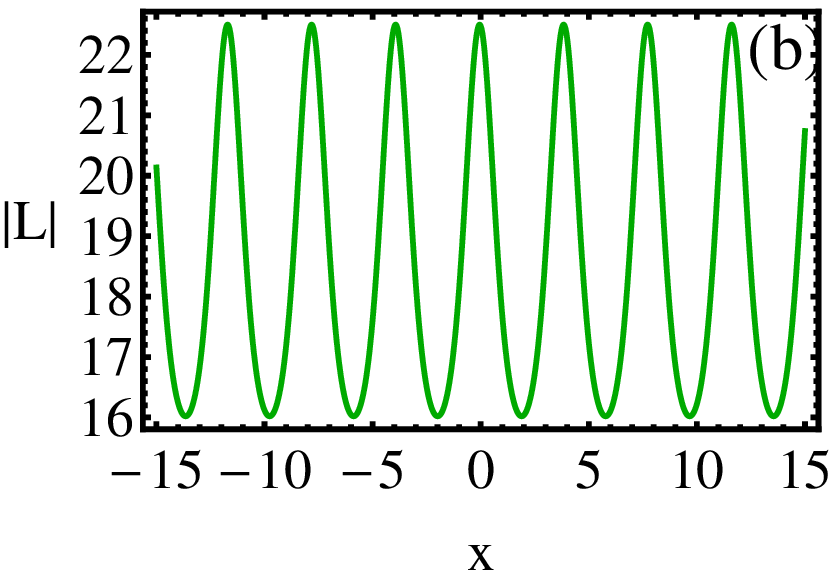}\\
\centering \includegraphics[height=4cm,width=5.8cm]{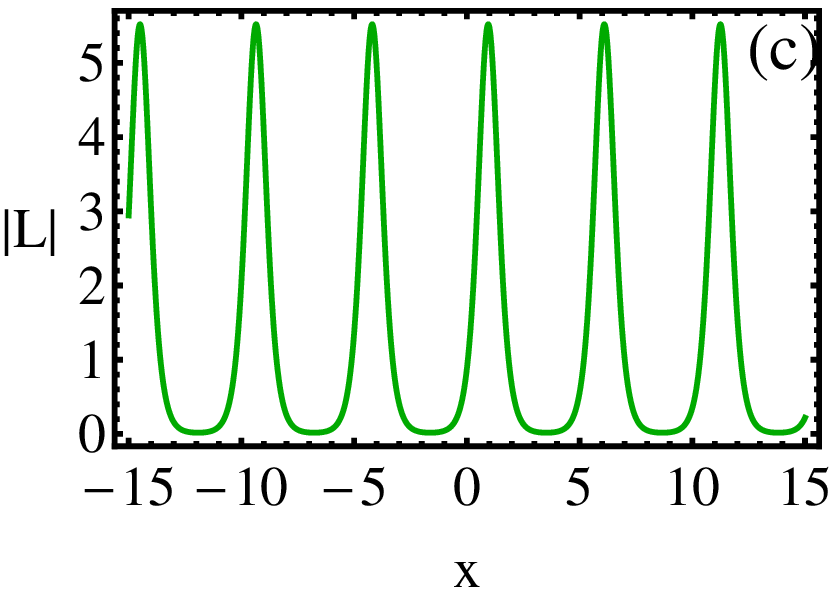}~~~\includegraphics[height=4cm,width=5.8cm]{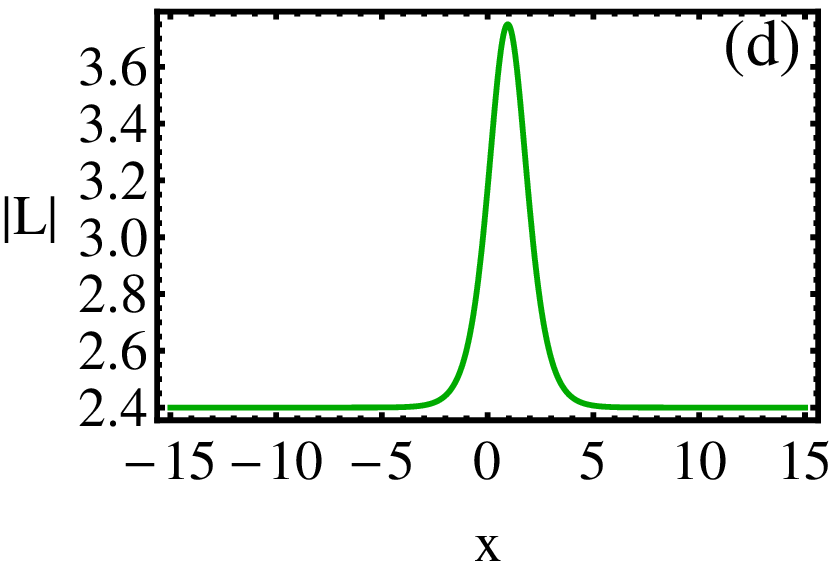}
\caption{Plots of long wave component corresponding to the short wave components given in Fig.~2.}
\end{figure}
In Fig.~2, we have plotted the profiles of the long wave corresponding to the short waves shown in Fig.~1, in a clockwise manner. In Fig.~2((a)-(c)), one can observe periodic solitary waves. A localized anti-dark solitary wave structure is depicted in Fig.~2(d). This anti-dark structure results from the boundary condition $L\longrightarrow$ constant as $x\longrightarrow \pm \infty$. Such anti-dark solitons has been observed in (2+1) dimensional generalized nonlinear Schr\"odinger equation \cite{nist} and in cubic-quintic nonlinear Scr\"odinger equation \cite{cros}. All the parameters have the same values as in Fig.~1.
\subsection{Solutions in terms of Lam\'e Polynomials of Order 2}
In this sub-section, we present the periodic solutions of order 2 (l=2) of the (2+1) dimensional LSRI system (\ref{2}). This case does not admit similar elliptic solutions. Here we obtain only mixed and superposed elliptic solutions which we present one by one.
\subsubsection{Mixed elliptic solutions}
We find there exist seven possible combinations of mixed elliptic solutions which are given below in Table 3.
\renewcommand{\floatpagefraction}{0.7}
\renewcommand{\arraystretch}{0.35}% Tighter
\begin{table}[H]
\caption {Mixed elliptic solutions of order 2}
\centering
\begin{tabular}{|c|c|c|l|l| }
  \hline
     Mixed&$f_{1}(u)$& $f_{2}(u)$ & Constraints on & Long wave\\\cline{2-3}
   elliptic&$S_{1}=f_{1}$&$S_{2}=f_{2}\times$&parameters& component\\
   solutions&$\times e^{-i(\omega_{1}t+\nu_{1} y-k_{1}x+\delta_{1})}$&$\times e^{-i(\omega_{2}t+\nu_{2} y-k_{2}x+\delta_{2})}$&~&~\\\hline
    \multirow{4}{1cm}{(1)} & \multirow{4}{2.5cm}{$A~\sqrt{m}~\textit{\cn (u,m)}$\\ $\times \textit{\dn (u,m)}~ $}& \multirow{4}{2.5cm}{$B~\sqrt{m}~\textit{\sn (u,m)}$ \\$\times \textit{\dn (u,m)} $}
          & $v+ \varepsilon_{1} w =2\delta k_{1}$,&~ \\
        & & & $~v + \varepsilon_{2} w =2\delta k_{2}$,&~\\
        & & & $~c_1 A^2 = c_2 B^2$,&~\\
        & & & $\beta^2 =  -\delta\frac{\left(m c_1 A^2\right)}{3v}>0$,&$\left(6 \delta \beta^{2}\right)\dn^{2}(u,m)$\\
        & & & $\omega_{1} +\varepsilon_{1} \nu_{1}-\delta\left(k_{1}^2 -(5-m) \beta^2\right) =0$,&~ \\
        & & & $\omega_{2} +\varepsilon_{2} \nu_{2}$&~\\
        & & &$-\delta\left(k_{2}^2 -(5-4m) \beta^2\right) = 0$.&~ \\
    \hline
    \end{tabular}
     \end{table}
\renewcommand{\floatpagefraction}{0.7}
\renewcommand{\arraystretch}{0.15}% Tighter
\begin{table}[H]
\centering
    \begin{tabular}{|c|c|c|c|c|}
    \hline
    \multirow{4}{1cm}{(2)} & \multirow{4}{2.5cm}{$A~\sqrt{m}~\textit{\cn (u,m)}$\\ $\times \textit{\dn (u,m)}~ $}& \multirow{4}{2.5cm}{$B~m~\textit{\sn (u,m)}$\\ $\times \textit{\cn (u,m)} $}
          & $v+ \varepsilon_{1} w =2\delta k_{1}$,&~ \\
        & & & $~v+ \varepsilon_{2} w =2\delta k_{2}$,&~\\
        & & & $~c_1 A^2 = c_2 B^2$,&~\\
        & & & $ \beta^2 =-\delta \frac{\left(c_1 A^2\right)}{3v}>0$,&$\left(6 m \delta \beta^{2}\right)~\cn^{2}(u,m)$ \\
        & & & $\omega_{1} + \varepsilon_{1} \nu_{1}$&~ \\
        & & & $-\delta(k_{1}^2-(5m-1) \beta^2) = 0$,&~\\
        & & & $\omega_{2} +\varepsilon_{1} \nu_{2}$&~ \\
        & & &$-\delta \left(k_{2}^2 -(5m-4) \beta^2\right) =0$.&~\\
    \hline
      \multirow{4}{1cm}{(3)} & \multirow{4}{2.5cm}{$A~\textit{\dn}^2 \textit{(u,m)}+D$}& \multirow{4}{2.5cm}{$B~m~\textit{\cn (u,m)}$\\$\times \textit{\sn (u,m)} $}
          & $v+ \varepsilon_{1} w =2\delta k_{1}$,&~ \\
          & & &$~v+ \varepsilon_{2} w =2\delta k_{2}$,&~\\
          & & &$~c_1 A^2 = c_2 B^2$,&$-2[\frac{c_{1}A^{2}(p+1)^{2}}{v}$\\
        & & & $\omega_{1} + \varepsilon_{1} \nu_{1}-\delta(k_{1}^2 -\big [5-4m+9p$&~\\
        & & &$+\frac{3m(1+p)}{2p+2-m} \big ] \beta^2)= 0$,&$+3 \delta \beta^{2}m\sn^{2}(u,m)]$  \\
        & & & $\beta^2 =-\delta\frac{(2p+2-m)c_1 A^2}{3v}>0$,&~\\
        & & & $\omega_{2} +\varepsilon_{2} \nu_{2}-\delta(k_{2}^2 -\big [3p-1-m $&~\\
        & & &$+\frac{3m(1+p)}{2p+2-m} \big ] \beta^2) =0$.&~ \\
    \hline
    \multirow{4}{1cm}{(4)} & \multirow{4}{2.5cm}{$A~\textit{\dn}^2 \textit{(u,m)}+D$}& \multirow{4}{2.5cm}{$B~\sqrt{m}~\textit{\sn (u,m)}$\\ $\times \textit{\dn (u,m)} $}
          & $v+ \varepsilon_{1} w =2\delta k_{1}$,&~ \\
        & & &$~v+ \varepsilon_{2} w =2\delta k_{2}$,&~\\
        & & &$~\beta^2 = -\delta\frac{\left((2p+1) c_1 A^2\right)}{3v}>0$,&~\\
        & & & $c_1 A^2 = c_2 B^2$,&$\left(6 \delta \beta^{2}\right)(\dn^{2}(u,m))$ \\
        & & & $\omega_{1} + \varepsilon_{1} \nu_{1}-\delta(k_{1}^2 -\big [4(2-m)+9p $&~\\
        & & &$-\frac{3p}{2p+1} \big ] \beta^2) =0$,&$-\frac{2}{v}c_{1} A^{2} p^{2}$ \\
        & & & $ \omega_{2} +\varepsilon_{2} \nu_{2}-\delta(k_{2}^2 -\big [5-4m+3p $&~\\
        & & &$-\frac{3p}{2p+1} \big ] \beta^2) =0$.&~ \\
    \hline
     \multirow{4}{1cm}{(5)} & \multirow{4}{2.5cm}{$A~\textit{\dn}^2 \textit{(u,m)}+D$}& \multirow{4}{2.5cm}{$B~\sqrt{m}~\textit{\cn(u,m)}$\\ $\times \textit{\dn (u,m)} $}
            & $v+ \varepsilon_{1} w =2\delta k_{1}$&~\\
          & & &$v+ \varepsilon_{2} w =2\delta k_{2}$&~\\
          & & &$~c_1 A^2 = -c_2 B^2$,&$\left(6 \delta \beta^{2}\right)(\dn^{2}(u,m))$\\
          & & & $~\beta^2 =-\delta\frac{\left((2p+1-m)c_1 A^2\right)}{3\nu}>0$&$-\frac{2p^{2}c_{1}A^{2}}{v}$\\
          & & & $\omega_{1} + \varepsilon_{1} \nu_{1}-\delta(k_{1}^2 -[4(2-m)+9p$&~\\
          & & &$ -\frac{3p(1-m)}{2p+1-m}]\beta^2)=0$,&~ \\
        & & & $\omega_{2}+\varepsilon_{2} \nu_{2}-\delta(k_{2}^2 -[5-m+3p $&~\\
        & & &$-\frac{3p(1-m)}{2p+1-m}]\beta^2) =0$.&~ \\
    \hline
    \multirow{4}{1cm}{(6)} & \multirow{4}{2.5cm}{$A~\sqrt{m}~\textit{\sn (u,m)}$\\$\times \textit{\dn (u,m)}$}& \multirow{4}{2.5cm}{$B~m~\textit{\cn (u,m)}$\\ $\times \textit{\sn (u,m)} $}
          & $\nu+ \varepsilon w =2\delta k_{1}$,&~\\
         & & &$~\nu+ \varepsilon w =2\delta k_{2}$&~\\
         & & &$~c_1 A^2 =- c_2 B^2$&$-6 \delta \beta^{2}m\sn^{2}(u,m)$ \\
        & & & $\beta^2 = \delta\frac{\left((1-m) c_1 A^2\right)}{3v}>0,$&~ \\
        & & & $\omega_{1} + \varepsilon_{1} \nu_{1}
-\delta\left(k_{1}^2 +(1+4m) \beta^2\right) =0$,&~ \\
        & & & $\omega_{2} +\varepsilon_{2} \nu_{2}
-\delta\left(k_{2}^2 +(4+m) \beta^2\right) =0$.&~ \\
    \hline
    \multirow{4}{1cm}{(7)} & \multirow{4}{2.5cm}{$A~\textit{\dn}^2 \textit{(u,m)}+D $}& \multirow{4}{2.5cm}{$B~\textit{\dn}^2 \textit{(u,m)}+E $}
          & $v+ \varepsilon_{1} w =2\delta k_{1}$,&$\left(3 \delta \beta^{2}\right)[2\dn^{2}(u,m)$\\
        & & &$~v+ \varepsilon_{2} w =2\delta k_{2}$,&$+\left(p+q\right)]$\\
        & & &$~c_1 A^2 = -c_2 B^2$,&~\\
        & & & $\beta^2 =  -\delta\frac{(2\left(p-q\right) c_1 A^2}{3v}>0$,&~ \\
        & & & $\omega_{1} + \varepsilon_{1}\nu_{1}-\delta\left(k_{1}^2 \mp 2 \sqrt{1-m+m^2} \beta^2\right) =0$,&~ \\
        & & & $\omega_{2} +\varepsilon_{2} \nu_{2}-\delta\left(k_{2}^2 \pm 2 \sqrt{1-m+m^2} \beta^2\right) =0$.&~ \\
        \hline
     \end{tabular}
   \end{table}
   In the above table, $p=\frac{D}{A}$,~$q=\frac{E}{B}$,~$p\neq q=-\frac{1}{3}\left[\left(2-m\right)\pm\sqrt{1-m+m^{2}}\right]$ and $(u,m)=(\beta(x-vt-wy+\delta_{0}),m)$. The above solution can be completely determined by the parameters $A$, $m$, $k_{1}$, $k_{2}$, $\nu_{1}$, $\nu_{2}$, $\delta_{0}$, $\delta_{1}$ and $\delta_{2}$. Note that in mixed elliptic solution (7), $p~ (=\frac{D}{A})$ and $q~(=\frac{E}{B})$ are different and hence the components $S_{1}$ and $S_{2}$ admit different intensity profiles. Unlike in the first order solutions, here we find that there is no intensity sharing among the components. This is quite clear from the expression for $\beta^{2}$, in the third column in Table 3, which is now expressed only in terms of A, not by the combination of A and B.
\subsubsection{Superposed elliptic solution}
The superposed elliptic solution of Eq. (\ref{2}) involving $\dn^{2}(u,m)$ and the product $\cn (u,m)\dn(u,m)$ is found to be
\bes \label{12}\bea
S_{1}(x,t)&=&\left[\frac{A}{2}~\textit{\dn}^2 \textit{(u,m)}~+D+\frac{F}{2} \sqrt{m}~\textit{\cn (u,m)}~\textit{\dn (u,m)}\right]~e^{-i(\omega_{1}t+\nu_{1} y-k_{1}x+\delta_{1})},\\
S_{2}(x,t)&=&\left[B\textit{\dn}^2 \textit{(u,m)}~+E+\frac{G}{2}~\sqrt{m}~\textit{\cn (u,m)}~\textit{\dn (u,m)}\right]~e^{-i(\omega_{2}t+\nu_{2} y-k_{2}x+\delta_{2})},\\
L&=&\left(3 \delta \beta^{2}\right)[(\dn^{2}(u,m)~\pm\sqrt{m}\cn(u,m)\dn(u,m))^{2}+p+q]\label{12c}.
\eea
The  validity of the solution requires\\
\bea
v+ \varepsilon_{1} w =2\delta k_{1},~v+ \varepsilon_{2} w =2\delta k_{2},~F=\pm A,~G=\pm B,~c_1 A^2 = -c_2 B^2,\nonumber\\~\beta^2 = -\delta \frac{\left(2(p -q)c_1 A^2\right)}{3\nu}>0,~p=\frac{D}{A},~q=\frac{E}{B},
~p\neq q=-\frac{1}{12} [(5-m) \pm \sqrt{1+14m+m^2}],\nonumber\\
~\omega_{1}+\varepsilon \nu_{1}-\delta\left(k_{1}^2 -[5-m+ 3(3p+q)]\beta^2\right) = 0,~ \omega_{2}+\varepsilon \nu_{2}-\delta\left(k_{2}^2-[5-m+ 3(p+3q)]\beta^2\right) = 0.
 \eea \ees
 Note that the signs of $F$ and $G$ are correlated, i.e. if $F=A~(F=-A)$ then $G=B~(G=-B)$.
\subsubsection{Hyperbolic solutions $m=1$}
 The interesting hyperbolic solutions can be deduced by fixing the modulus parameter $m=1$ in the above second order elliptic function solutions.\\
(a) The mixed solutions (1) and (2) of order 2 given in Table 3, reduce to the following hyperbolic form
\bes \label{13}\bea\label{13a}
S_{1}&=&\left[A~\sech^{2}(\beta(x-v t-w y+\delta_{0}))\right]~e^{-i(\omega_{1}t+\nu_{1} y-k_{1}x+\delta_{1})},\\\label{13b}
S_{2}&=&\left[B~\sech(\beta(x-v t-w y+\delta_{0}))~\tanh(\beta(x-vt-wy+\delta_{0}))\right]~e^{-i(\omega_{2}t+\nu_{2} y-k_{2}x+\delta_{2})},\\
L&=&\left(6 \delta \beta^{2}\right) \sech^{2}(\beta(x-vt-wy+\delta_{0}))\label{13c},
\eea
along with the parametric restriction\\
\bea
v+ \varepsilon_{1}w =2\delta k_{1},~v+ \varepsilon_{2} w =2\delta k_{2},~c_1 A^2 = c_2 B^2,~\beta^2 = -\delta\frac{c_1 A^2}{3v}>0,\nonumber\\
\omega_{1}+\varepsilon_{1} \nu _{1}-\delta\left(k_{1}^2-4\beta^2\right) = 0,~ \omega_{2}+\varepsilon_{2} \nu _{2}-\delta\left(k_{2}^2-\beta^2\right)= 0.
\eea \ees
The above solitary wave solutions can be viewed as the blue-white-blue solutions reported in Ref.~\cite{hioe1} for the three coupled nonlinear Schr\"odinger system.\\
(b) It can be easily verified that the mixed solutions (3) and (4) of order 2 given in Table 3, either go over to above solutions (\ref{13}) with $D=0$ or reduce to the hyperbolic solution
 \bes \label{14}\bea
S_{1}&=&\left[A~\sech^{2}(\beta(x-vt-wy+\delta_{0}))~+D\right]~e^{-i(\omega_{1}t+\nu_{1} y-k_{1}x+\delta_{1})},\label{14a}\\
S_{2}&=&\left[B~\sech(\beta(x-vt-wy+\delta_{0}))~\tanh(\beta(x-vt-wy+\delta_{0}))\right]~e^{-i(\omega_{2}t+\nu_{2} y-k_{2}x+\delta_{2})},\label{14b}\\
L&=&\left(6 \delta \beta ^{2}\right) \sech^{2}(\beta(x-vt-wy+\delta_{0}))-8~\delta~\beta^{2}\label{14c},
\eea
along with the parametric restriction\\
\bea\label{14d}
v + \varepsilon_{1}w =2\delta k_{1},~v + \varepsilon_{2} w =2\delta k_{2},~
~c_1 A^2 = c_2 B^2,~\beta^2 =\delta\frac{\left(c_1 A^2\right)}{9v}>0,\nonumber\\
\omega_{1}+\varepsilon_{1} \nu_{1} -\delta\left(k_{1}^2 +8\beta^2\right) = 0,~\frac{D}{A}=p=-\frac{2}{3},~ \omega_{2}+\varepsilon_{2} \nu_{2}-\delta\left(k_{2}^2+7\beta^2\right)= 0.
\eea \ees
 The above solitary wave solutions are similar to  the red-white-red solutions discussed in Ref.~\cite{hioe1} for the three coupled nonlinear Schr\"odinger system. But these solution behave different from that of Ref.~\cite{hioe1} due to the constraint condition (\ref{14d}).\\

 (c) The mixed elliptic solution (5) given in Table 3 takes the hyperbolic form given below:
 \bes \label{15}\bea
S_{1}&=&\left[A~\sech^{2}(\beta(x-vt-u y+\delta_{0}))~+D\right]~e^{-i(\omega_{1}t+\nu_{1} y-k_{1}x+\delta_{1})},\label{15a}\\
S_{2}&=&\left[B~\sech^{2}(\beta(x-vt-u y+\delta_{0}))\right]~e^{-i(\omega_{2}t+\nu_{2}y-k_{2}x+\delta_{2})},\label{15b}\\
L&=&-\left(9 \delta \beta ^{2}\right)\left[ p~\sech^{2}(\beta(x-vt-wy+\delta_{0}))+\frac{p^{2}}{2}\right]\label{15c},
\eea
where the parameters are constrained by the following conditions: \\
\bea
v + \varepsilon_{1}w =2\delta k_{1},~v+ \varepsilon_{2}w =2\delta k_{2},~\
~c_1 A^2 = -c_2 B^2,~\beta^2 =\delta\frac{\left(4c_1 A^2\right)}{9v}>0,\nonumber\\
\omega_{1}+\varepsilon_{1} \nu_{1} -\delta\left(k_{1}^2 +2\beta^2\right) = 0,~\frac{D}{A}=p=-\frac{2}{3},~ \omega_{2}+\varepsilon_{2}\nu_{2}-\delta\left(k_{2}^2-2\beta^2\right)= 0.
\eea \ees
This solution looks like red-blue-red solutions discussed in Ref.~\cite{hioe1}.
 In the limit $m=1$, mixed elliptic solution (7) given in Table 3 and the superposed solution (\ref{12}) go over to solution (\ref{15}), i.e., for these solutions $D\neq 0$, $E=0$ and rest of the solution is as given by (\ref{15}). Needless to say, there is another solution similar to (\ref{15}), i.e., in that case $E\neq 0$, $D=0$.  Finally, in the limit $m=1$, the mixed elliptic solution (6) in Table 3 does not exist as $\beta^{2}$ becomes zero for this choice.
\renewcommand{\floatpagefraction}{0.7}
\begin{figure}[H]
\centering\includegraphics[height=4cm,width=5.5cm]{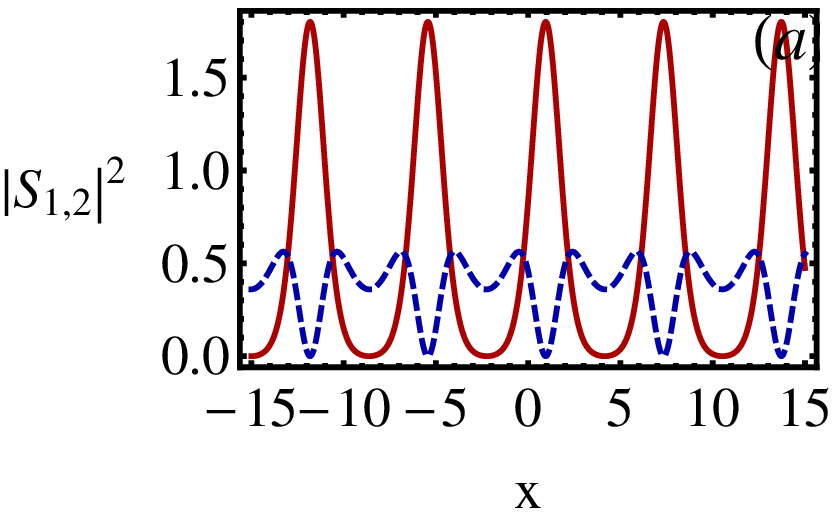}~\includegraphics[height=4cm,width=5.5cm]{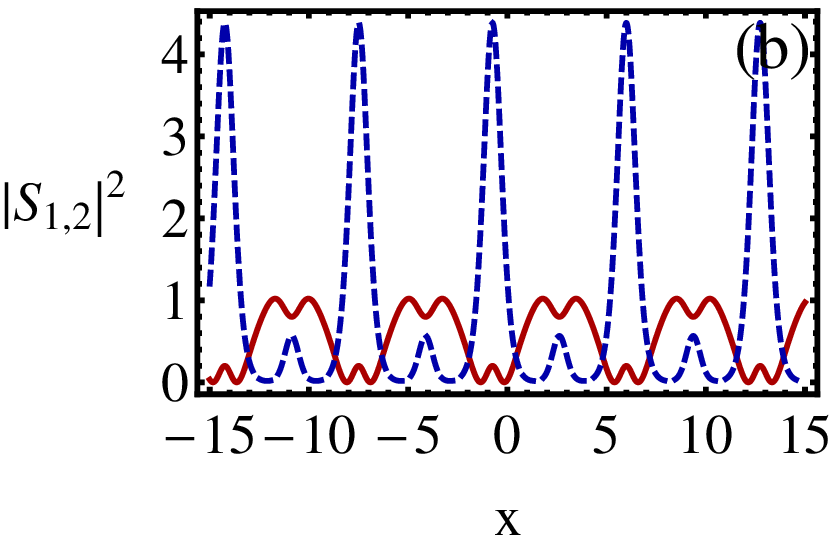}~
\includegraphics[height=4cm,width=5.5cm]{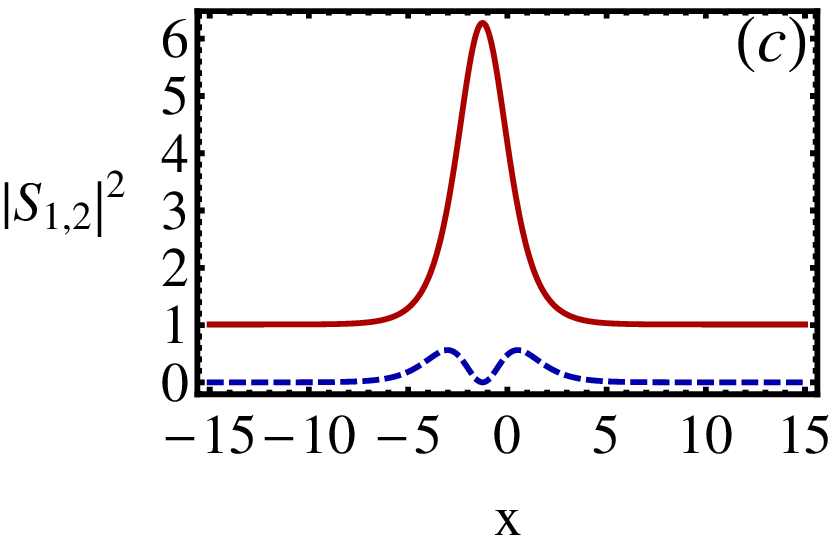}
\caption{ Red and dashed blue plots correspond to first and second components of short wave, respectively. (a) intensity plot of mixed elliptic solution  (1) given in Table 3, for the choice $v=1.2$, $m=0.8$. (b) superposed solution given by Eq. (\ref{12}) for the choice $v=-0.5$, $m=0.8$ and for the condition $c_{1}=-c_{2}=1$. (c) intensity plot of anti-dark solitary wave solution in $S_1$ component and non-trivial bright bi-solitary wave solution in $S_2$ component given by Eqs. (\ref{14a}) and (\ref{14b}) for the choice $A=B$, $v=-1.0$, $m=1$. For Fig.~3(a) and Fig.~3(c) $c_{1}=c_{2}=1$. In all the figures, $y=0.5$, $t=1$, the parameters $A=1.5$, $\nu_{1}=\nu_{2}=1.2$, $w=1.5$,  $\delta=-1$, $\varepsilon_{1,2}=1$, and $\delta_{0}=\delta_{1}=\delta_{2}=1$. In the case of $\delta=1$ also, we get same thing, so simply we present the $\delta=-1$ case only.}
\end{figure}
For illustrative purpose we have plotted some of the second order solutions in Fig.~3. In Fig.~3(a) the  periodic waves given by mixed elliptic solution (1) in Table 3 is plotted. Here $S_{1}$ admits periodic solitary wave train and the $S_{2}$ admits doubly-periodic wave train. The superposed periodic waves given by Eq. (\ref{12}) are shown in Fig.~3(b). These superposed solutions show significant amplification in the pulse train. Finally, we have depicted the hyperbolic solutions given by Eqs. (\ref{14a}) and (\ref{14b}) in Fig.3(c). Here the first component ($S_1$) admits anti-dark solitary wave profile and the second component ($S_2$) has a bi-solitary wave profile.
\renewcommand{\floatpagefraction}{0.7}
\begin{figure}[H]
\centering\includegraphics[height=4cm,width=5.1cm]{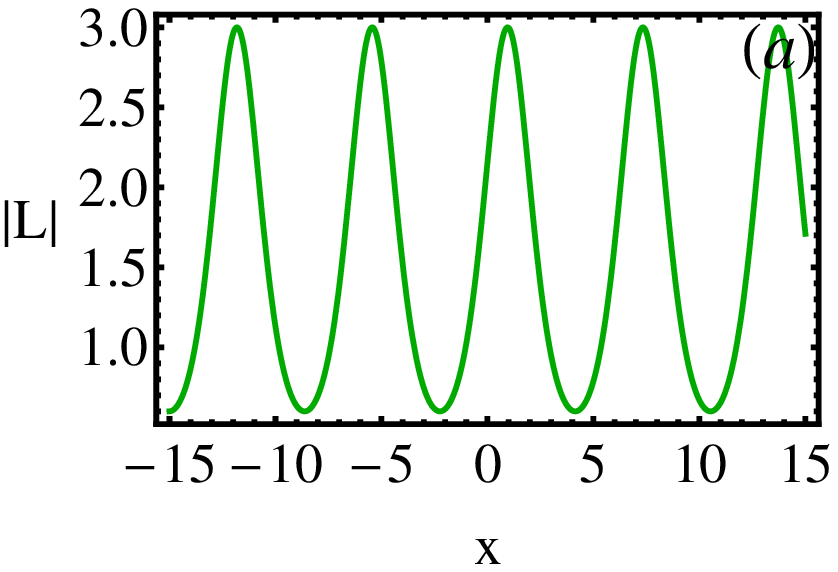}~~~~\includegraphics[height=4cm,width=5.1cm]{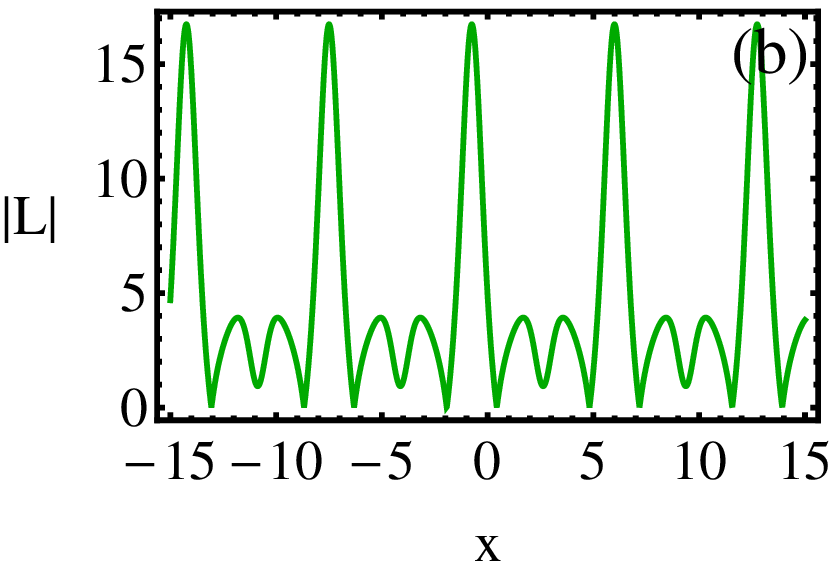}~~~~
\includegraphics[height=4cm,width=5.1cm]{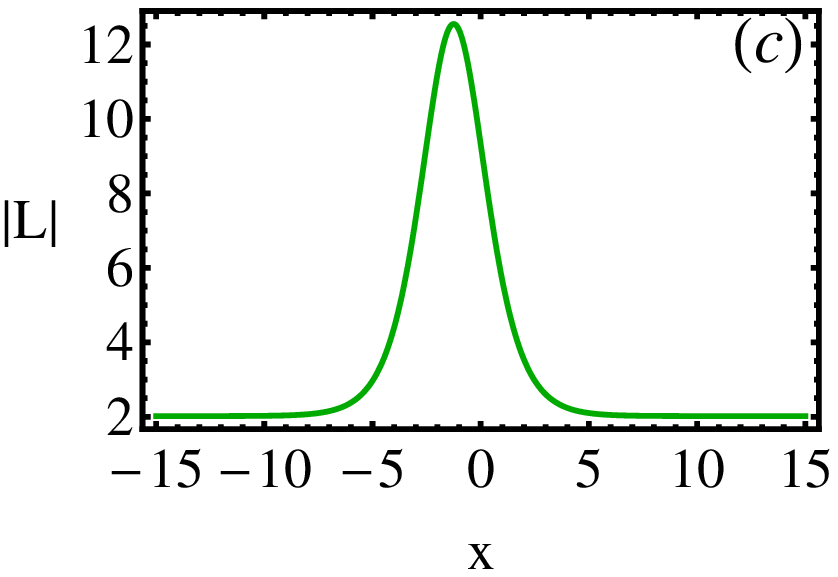}
\caption{  Intensity plots of long wave component corresponding to the short wave components given in Fig.~(4).}
\end{figure}
Fig.~4(a) displays periodic solitary wave trains in the LW component (see solution 1 in Table 3) while Fig.~4(b) shows a doubly periodic wave train given by Eq. (\ref{13c}). Fig.~4(c) shows an anti-dark solitary wave like structure given by Eq. (\ref{14c}). One can verify that (\ref{15c}) also admits such anti-dark solitary wave.
\section{Jacobi Elliptic function solutions of the (1+1)-Dimensional two component LSRI system}
In this section, we construct the Jacobi elliptic solution for (1+1) dimensional two component LSRI system (1). We can follow the same mathematical treatment as in the case of (2+1) dimensional case. Here also, we choose the travelling wave ansatz.
\bea\label{16}
S_{j}=f_{j} [\beta(x-v t+\delta_{0})]e^{-i(\omega_{j} t-k x+\delta_{j})}, \quad\quad j=1,2,
\eea
to construct the elliptic function solution of Eq. (\ref{1a}). Here $f_{j}$ are   real functions of $x$ and $t$; $\beta$, $\delta_0$ and $\delta_{1,2}$ are  arbitrary real constants, $ \omega_{j} $ is the frequency of the $j^{th}$ SW component, $k$ is the wave number and $v$ is the velocity of the short wave. Note that both the short waves are travelling with same velocity.

By substituting the above ansatz (\ref{16}) into Eq. (\ref{1b}), we obtain  the LW component as
\bea\label{17}
L=\frac{-2}{v}(c_{1}|S_{1}|^2+c_{2}|S_{2}|^2).
\eea
Substitution of  the ansatz (\ref{16}) in (\ref{1a})  and using (\ref{17}) result in a set of complex equations. On equating real and imaginary parts of the complex equations we respectively get
\bes\bea
\frac{d^2 f_j}{du^2}+\left[ \frac{\delta \omega_j-k^{2}}{\beta^2}-\frac{2\delta}{v \beta^2}\left(c_1 f_1^2+c_2 f_2^2\right)\right]f_j=0,\;\;\;j=1,2, \label{17a}\\
v=2\delta k.\qquad \qquad \qquad \qquad \qquad \qquad
\eea \ees
Here $\delta = \pm1$ and $u=\beta(x-vt+\delta_{0})$.
Next, we assume the Lam\'e function ansatz for $f_{j}$, that is,
\bea\label{18}
f_j=\rho_j \psi_{j}^{(l)},   \quad \quad l,j=1,2,
\eea
where $\psi_{j}^{(l)}$ is the first (second) order Lam\'e polynomials for $l=1$ ($l=2$), and satisfies the Lam\'e equation (\ref{7}).
\subsection{Solutions in terms of Lam\'e polynomials}
As in the (2+1) dimensional case, here also we obtain similar, mixed, superposed elliptic solution of order one and mixed as well as superposed solutions of order two. For brevity, we mention how to write these solutions from their (2+1) dimensional counterparts. The first order solutions for the similar and mixed cases will be the same as given in Tables 1 and 2 while the second order mixed solutions will be similar to those given in Table 3 with  $\varepsilon_{1}=\varepsilon_{2}=\nu_{1}=\nu_{2}=0$, $k_{1}=k_{2}=k$. Here one has to consider the corresponding constraint with this choice for a given pair of elliptic function solutions. For example, for this choice the similar elliptic solution (1) presented in Table 1 becomes;
\bes\bea
S_{1}&=&A~\dn(u,m) e^{-i (\omega_{1}-k x +\delta_{1})},\\
S_{2}&=&B~\dn(u,m) e^{-i (\omega_{2}-k x +\delta_{2})},\\
L&=&\left[\left(2 \delta  \beta^{2}\right)\dn^{2}(u,m)\right],
\eea
where $u=\beta(x-v t+\delta_{0})$, with the parametric restriction\\
\bea
v =2\delta k,~\beta^2 =-\delta\frac{\left(c_{1} A^{2}+c_{2} B^{2}\right)}{v}>0,~
\omega_{1}=\delta \left[k^{2} -(2-m)\beta^{2}\right],~ \omega_{2}=\omega_{1}.
\eea\ees
In a similar way all the first and second order solutions can be written. Similarly the first and second order superposed elliptic solutions and hyperbolic solutions also can be written from the corresponding two dimensional solutions by restricting the parameters as $\varepsilon_{1}=\varepsilon_{2}=\nu_{1}=\nu_{2}=0$, $k_{1}=k_{2}=k$. The associated constraint conditions also have to be rewritten appropriately with the above choice of parameters.
\begin{figure}[H]
\centering\includegraphics[height=4cm, width=5.9cm]{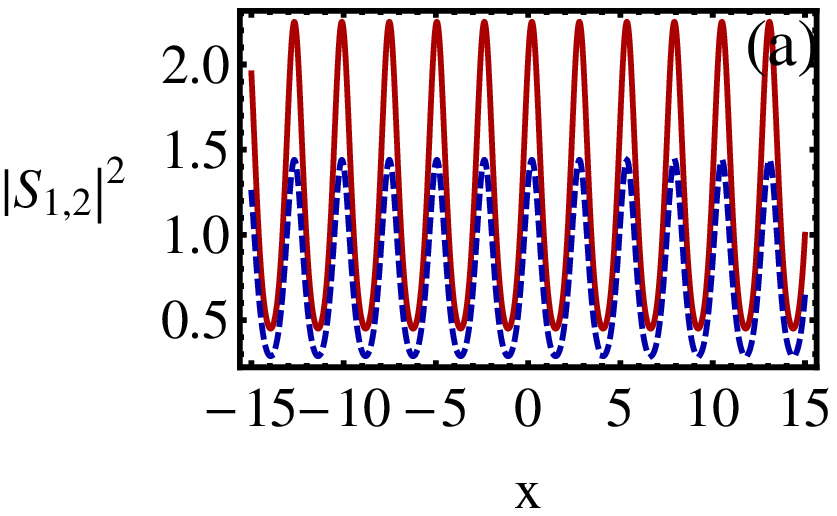}~~~~
\includegraphics[height=4cm,width=5.9cm]{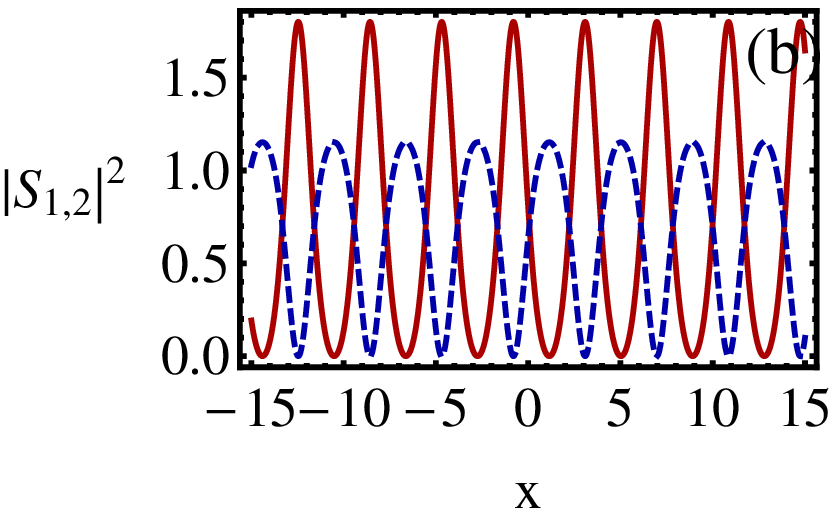}\\
\includegraphics[height=4cm,width=6cm]{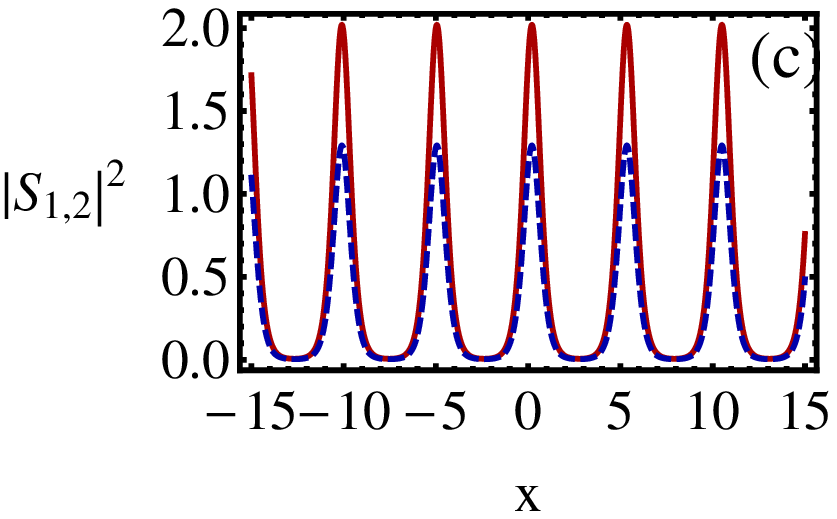}~~~~
\includegraphics[height=4cm,width=5.9cm]{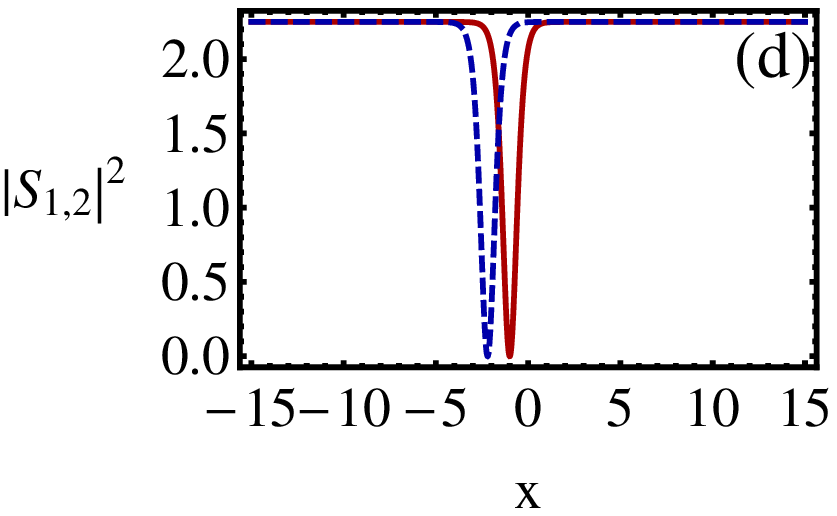}\\
\caption{Red and dashed blue plots correspond to first and second components of short wave, respectively. (a) intensity plot of similar elliptic solution (1) given in Table 1, for the velocity value  $v=1.2$, $m=0.8$. (b) intensity plot of mixed elliptic solution  (2) given in Table 2, for the velocity value $v=0.2$, $m=0.8$. (c) superposed solution given by Eq. (\ref{8}) for the velocity value $v=1.2$, $m=0.8$ and for the condition $A=D$, $B=E$. (d) plot of dark-dark solitary wave solution given by Eqs. (\ref{11a}) and (\ref{11b}) for the velocity value $A=1.2$, $B=1.2$, $v=1.2$, $m=1$. In all the figures, $y=0$, $t=1$, the parameters $A=1.5$ (except the Fig.~5(c)), $B=1.2$, $\nu_{1}=\nu_{2}=0$, $w=0$, $\delta=-1$, $\varepsilon_{1,2}=0$, $\delta_{0}=\delta_{1}=\delta_{2}=c_{1}=c_{2}=1$.}\label{Figure1}
\end{figure}
For illustrative purpose,we present some of the intensity plots of elliptic and hyperbolic solutions of the short wave components in Fig.~5. Here, Fig.~5(a) shows the periodic solution given by the similar solution (1) in Table 1, with the choice of parameters $\varepsilon_{1}=\varepsilon_{2}=\nu_{1}=\nu_{2}=0$, $k_{1}=k_{2}=k$. Fig.~5(b) shows the intensity plot of the mixed elliptic solution (2) given in Table 2, with the same above choice of parameters. It can be observed from the Fig.~5(a) and Fig.~5(b) that associated SW solutions are in-phase and out-of-phase. Then, superposed periodic wave solutions for the short wave components given by Eqs. (\ref{8a}) and (\ref{8b}) with the choice of parameters $\varepsilon_{1}=\varepsilon_{2}=\nu_{1}=\nu_{2}=0$, $k_{1}=k_{2}=k$ are plotted in Fig.~5(c). Finally, Fig.~5(d), depicts hyperbolic (dark-dark soliton) solution given by Eqs. (\ref{11a}) and (\ref{11b}) for the same above choice of parameters $\varepsilon_{1}=\varepsilon_{2}=\nu_{1}=\nu_{2}=0$, $k_{1}=k_{2}=k$. Here the first and second SW components are comprised of  dark solitary waves.
\renewcommand{\floatpagefraction}{0.7}
\begin{figure}[H]
\centering\includegraphics[height=3.8cm,width=5.1cm]{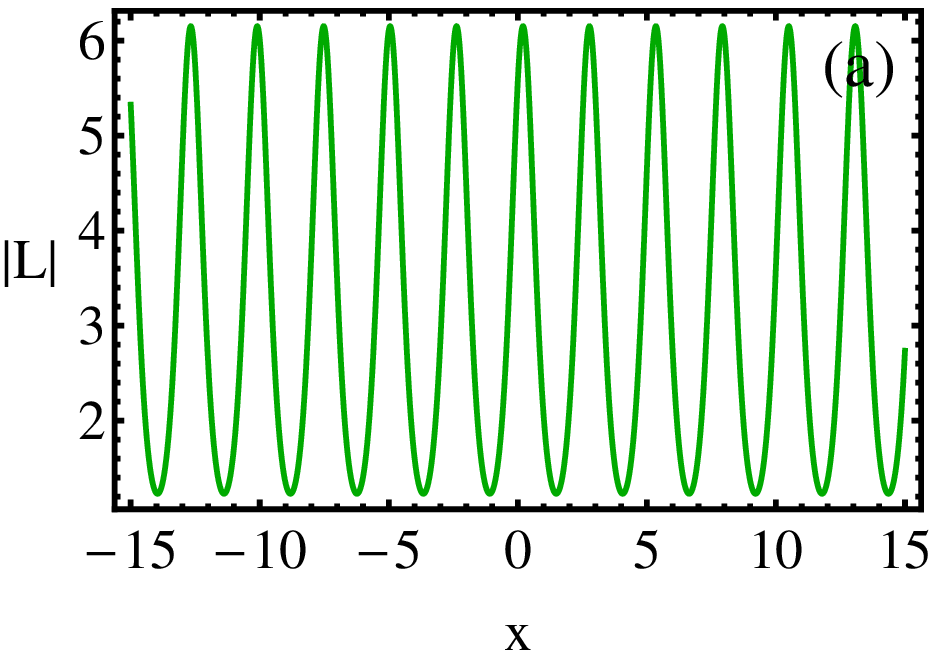}~~~~\includegraphics[height=3.9cm,width=5.1cm]{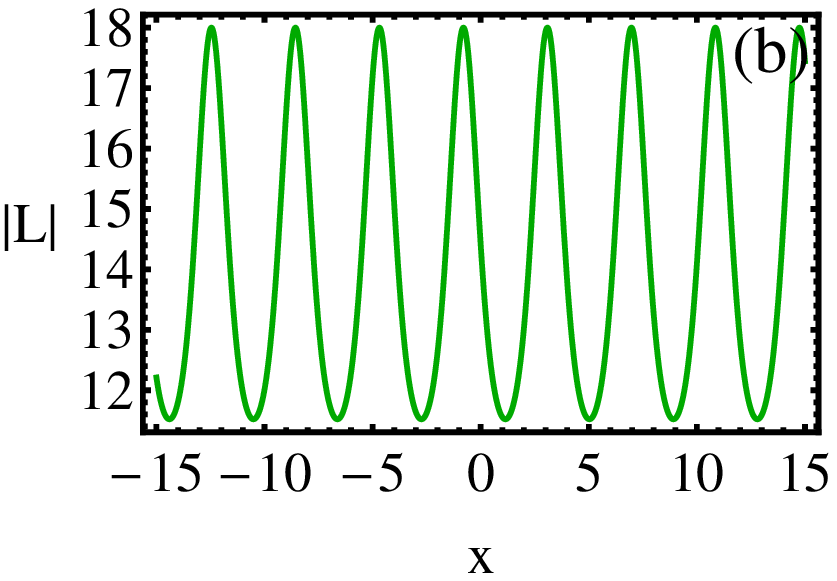}\\ \includegraphics[height=3.9cm,width=5.1cm]{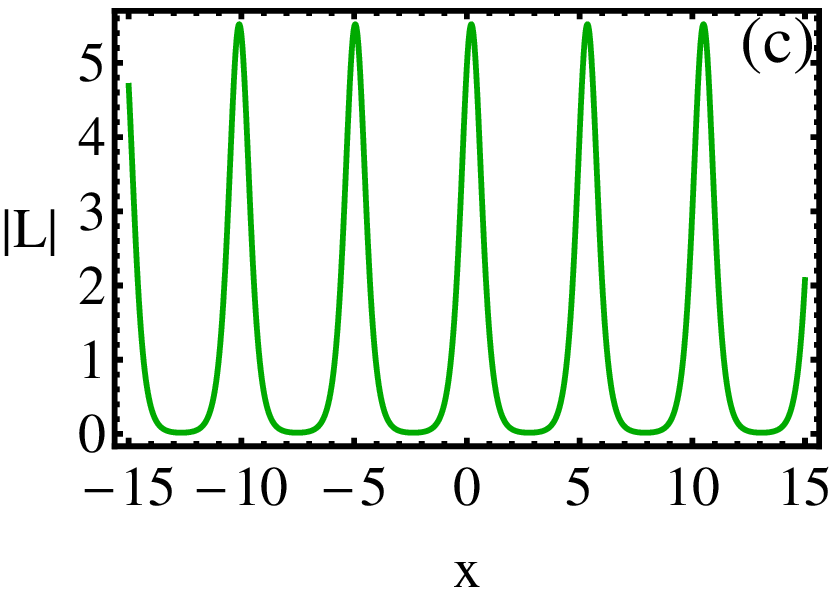}~~~~\includegraphics[height=3.9cm,width=5cm]{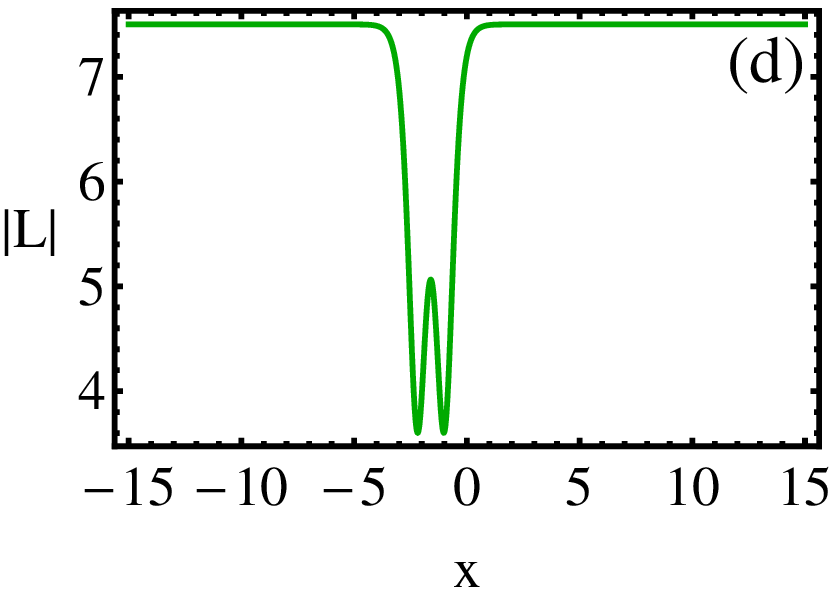}
\caption{Figures.~6((a)-(d)) depict the long wave component corresponding to the short wave components given in Figure.~5((a)-(d)), respectively.}\label{Figure2}
\end{figure}
 In Fig.~6, we have plotted the profiles of the long wave corresponding to the short waves shown in Fig.~5. All the arbitrary parameters are same as in Fig.~5. Interestingly, in Fig.~5(d) corresponding to the solution given by Eq. (\ref{11c}) with choice of parameters $\varepsilon_{1}=\varepsilon_{2}=\nu_{1}=\nu_{2}=0$, $k_{1}=k_{2}=k$, we observe a grey W type solitary wave.
%%%%%%%%%%%%%%%%%%%%%%%%%%%%
\renewcommand{\floatpagefraction}{0.7}
\begin{figure}[H]
\centering\includegraphics[height=4cm,width=5.2cm]{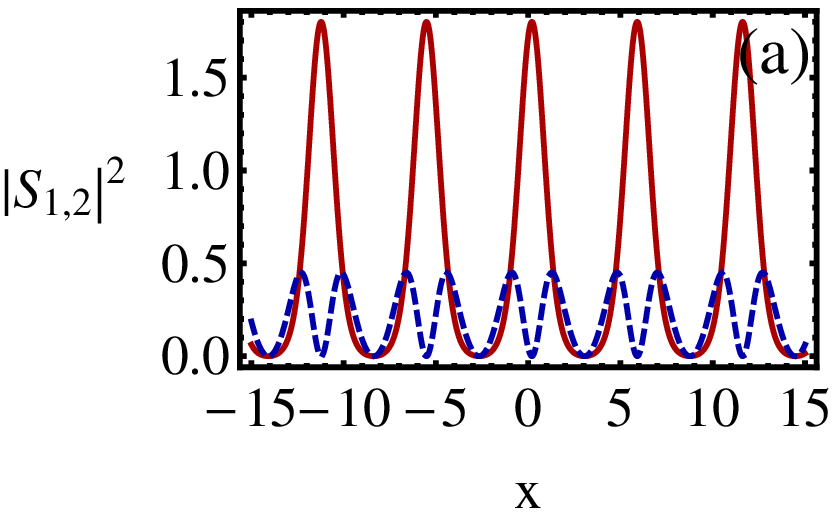}~~~\includegraphics[height=4cm,width=5.2cm]{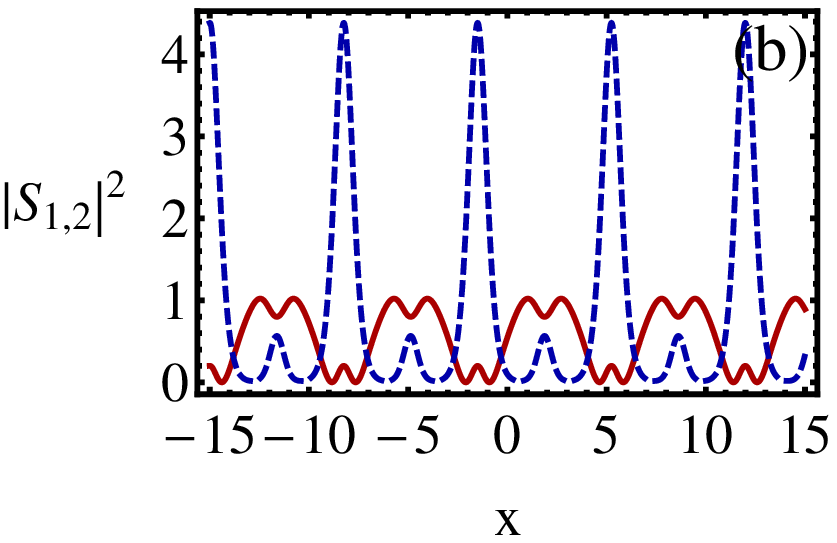}~~~
\includegraphics[height=4cm,width=5.2cm]{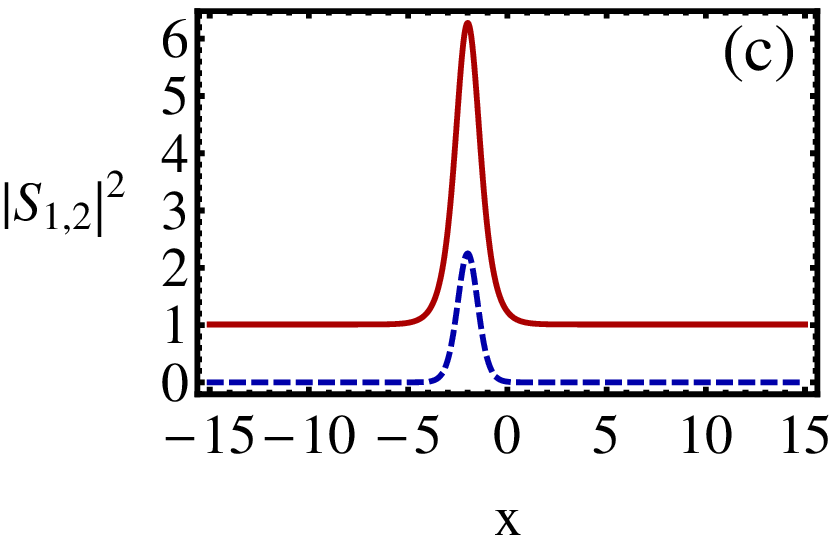}
\caption{ Red and dashed blue plots correspond to first and second components of short wave, respectively. (a) intensity plot of mixed elliptic solution  (2) given in Table 3, for the choice $v=1.2$, $m=0.8$. (b) superposed solution given by Eq. (\ref{12}) for the choice $v=-0.5$, $m=0.8$ and for the condition $c_{1}=-c_{2}=1$. (c) intensity plot of anti-dark solitary wave solution in $S_1$ component and regular bright-solitary wave solution in $S_2$ component given by Eqs. (\ref{15a}) and (\ref{15b}) for the choice $A=B$, $v=-1.0$, $m=1$. For Fig.~3(a) and Fig.~3(c) $c_{1}=c_{2}=1$. In all the figures, $y=0$, $t=1$, the parameters $A=1.5$, $\nu_{1}=\nu_{2}=0$, $w=0$,  $\delta=-1$, $\varepsilon_{1,2}=0$, and $\delta_{0}=\delta_{1}=\delta_{2}=1$. In the case of $\delta=1$ also, we get same thing, so simply we present the $\delta=-1$ case only.}
\end{figure}
 Fig.~7(a) shows the periodic waves given by the second order mixed solution (2) in Table 3 with the choice of parameters $\varepsilon_{1}=\varepsilon_{2}=\nu_{1}=\nu_{2}=0$, $k_{1}=k_{2}=k$. Here the periodic wave train for the second component $S_2$ is intricate and it attains one minimum intensity when the other component $S_1$ reaches the maximum, while it reaches the second minimum exactly at the minimum of the second component $S_2$. Fig.~7(b) depicts the superposed periodic wave given by Eq. (\ref{12}) with the same choice of above parameters, but with significant compression in their widths and amplification in their amplitudes. Finally, in Fig.~7(c) we have depicted the hyperbolic solutions given by Eq. (\ref{15}) with the choice of parameters $\varepsilon_{1}=\varepsilon_{2}=\nu_{1}=\nu_{2}=0$, $k_{1}=k_{2}=k$. Here the first component ($S_1$) is comprised of anti-dark solitary wave while  the second component ($S_2$) admits standard bright solitary wave.
\renewcommand{\floatpagefraction}{0.7}
\begin{figure}[H]
\centering\includegraphics[height=4cm,width=5.1cm]{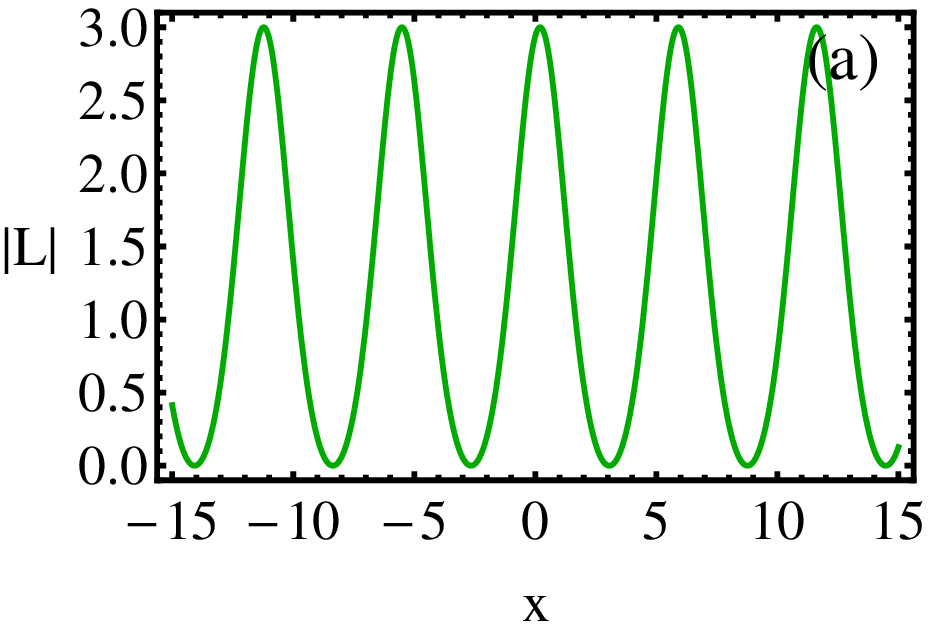}~~~\includegraphics[height=4cm,width=5.1cm]{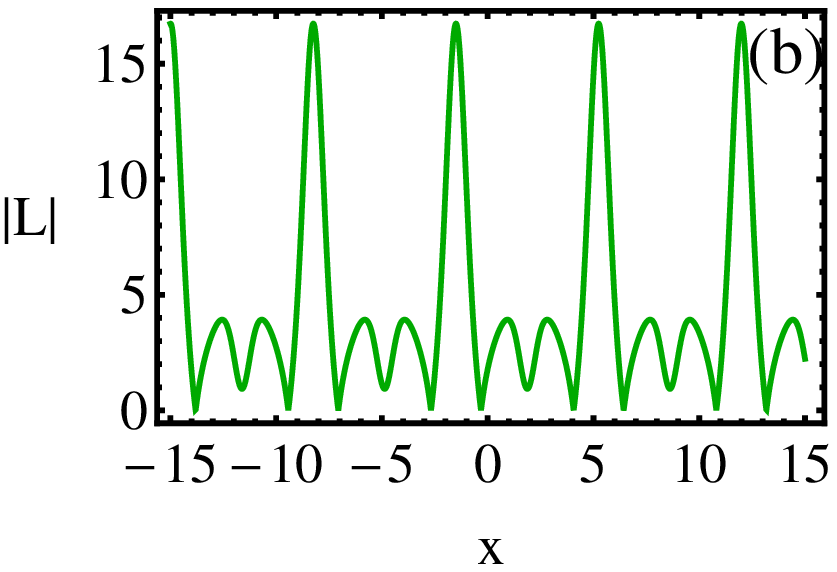}~~~
\includegraphics[height=4cm,width=5.1cm]{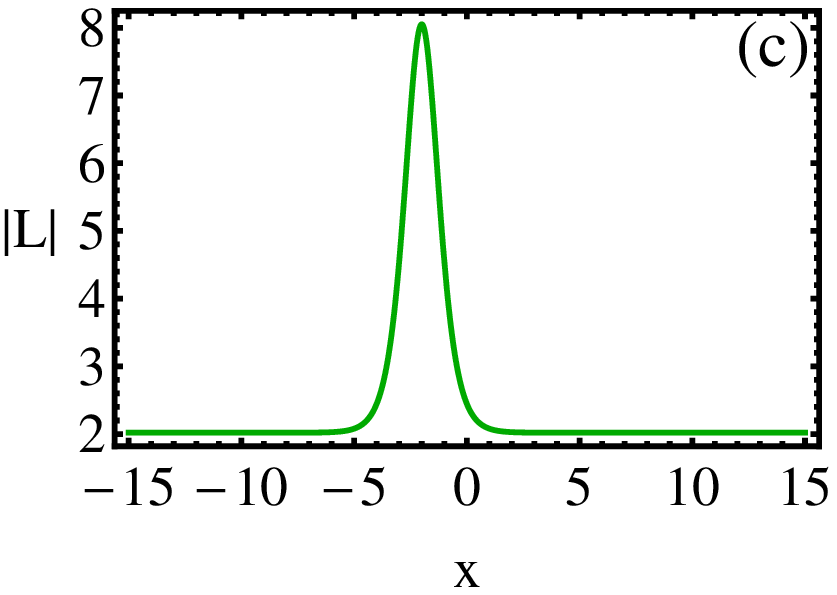}
\caption{Plots of long wave component corresponding to the short wave components given in Fig.~7.}\label{Figure2}
\end{figure}
 In Fig.~8, we have plotted the profile of the long wave corresponding to the short waves shown in Fig.~7. All the arbitrary parameters are same as in Fig.~7.
\section{Conclusion}
To conclude, first we obtain the Jacobi elliptic function solutions of two component (2+1) dimensional LSRI system (2) in terms of Lam\'e polynomials of order 1 and 2. We classified the solutions as similar, mixed and superposed elliptic waves based on their wave profile in the two SW components. We have reported special in-phase and out-of phase periodic wave train solutions. The similar elliptic solutions do not exist for the second order solutions. By considering the hyperbolic limit, we have identified  anti-dark and double-hump solitary waves in the SW component with an anti-dark solitary wave like structure in the LW component. Then we extend the same mathematical treatment to the (1+1) dimensional two component LSRI system and constructed order-1 and order-2 elliptic waves. We have demonstrated that the SW components as well as LW component support rich profile structures like bright, dark, anti-dark and grey W type solitary waves. It is of future interest to generalize this study to M-component LSRI system with $M>2$. There are several other coupled systems \cite{zuo}-\cite{tkpla} and higher dimensional systems \cite{zhen1} for which such elliptic wave solutions can be constructed. It will also be an interesting further work to numerically study the non-integrable versions of Eqs. (\ref{1}) and (\ref{2}) by making use of the elliptic function solutions as well as hyperbolic solutions reported here.
\section*{ACKNOWLEDGMENTS}
The author AK acknowledges Department of Atomic Energy (DAE). Govt. of India for the financial support through Raja Ramanna Fellowship. The authors TK and KT acknowledge the Principal and Management of Bishop Heber College, Tiruchirapalli, for constant support and encouragement.

\end{document}